\documentclass[sn-mathphys,iicol]{sn-jnl_mod}
\usepackage{amsmath}
\usepackage{xcolor}
\usepackage{amssymb}
\newcommand{\ket}[1]{\left\rvert#1\right>}
\newcommand{\bra}[1]{\left<#1\right\rvert}
\newcommand{\braket}[3]{\left<#1\right\rvert#2\left\rvert#3\right>}
\usepackage{hyperref}
\usepackage{amsfonts}       
\usepackage{upgreek}

\usepackage{array}
\usepackage{mathrsfs}
\usepackage{graphicx}
\usepackage{float}          
\usepackage{subcaption}
\usepackage{cuted}
    
\newenvironment{widetext}
    {\begin{strip}
     \rule{\dimexpr(0.5\textwidth-0.5\columnsep)}{0.4pt}\vrule height 6pt}
    {\par
    \hfill
    \rule[0.5\baselineskip]{\dimexpr(0.5\textwidth-0.5\columnsep-1pt)}{0.4pt}
    \par
    \vspace{-12.2pt}\hspace{0.5\textwidth}\hspace{9pt}
    \vrule height 6pt
      \end{strip}}
\newenvironment{widetext2}
    {\begin{strip}
     \rule{\dimexpr(0.5\textwidth-0.5\columnsep)}{0.4pt}\vrule height 6pt}
    {\end{strip}}

\jyear{2022}%

\theoremstyle{thmstyleone}%
\theoremstyle{thmstyletwo}%

\theoremstyle{thmstylethree}%

\begin{document}

\title[Aharonov-Bohm effect in phase space]{Aharonov-Bohm effect in phase space}

\author[1,2]{\fnm{Jose A. R.} \sur{Cembranos}}\email{cembra@ucm.es}

\author*[1]{\fnm{David} \sur{García-López}}\email{davgar20@ucm.es}

\author[1]{\fnm{Zoe} \sur{G. del Toro}}\email{zoe.garcia@tum.de}

\affil[1]{\orgdiv{Departamento de Física Teórica}, \orgname{Universidad Complutense de Madrid},\\ \orgaddress{\street{Plaza de Ciencias 1}, \city{Facultad de Ciencias Físicas}, \postcode{28040}, \state{Madrid},
\country{Spain}}}

\affil[2]{\orgdiv{Institute of Particle and Cosmos Physics (IPARCOS)}, \orgname{Universidad Complutense de Madrid}, \orgaddress{\street{Plaza de Ciencias 1}, \city{Facultad de Ciencias Físicas}, \postcode{28040}, \state{Madrid},
\country{Spain}}}



\abstract{
The Aharonov-Bohm effect is a genuine quantum effect typically characterized by a measurable phase shift in the wave function for a charged particle that encircles an electromagnetic field located in a region inaccessible to the mentioned particle. However, this definition is not possible in the majority of the phase space descriptions since they are based on quasiprobability distributions. In this work, we characterize for the first time the Aharonov-Bohm effect within two different formalisms of quantum mechanics. One of them is the phase-space formalism relying on the canonical commutation relations and Weyl transform. In this framework, the aim is to obtain a consistent description of the quantum system by means of the quasiprobability Wigner function. The other one is the Segal-Bargmann formalism, which we mathematically describe and connect with quantum mechanics by means of the commutation relations of the creation and annihilation operators. After an introduction of both formalisms, we study the Aharonov-Bohm effect within them for two specific cases: One determined by a non-zero electric potential, and another determined by a non-zero magnetic vector potential. Subsequently, we obtain a more general description of the Aharonov-Bohm effect that encompasses the two previous cases and that we prove to be equivalent to the well-known description of this effect in the usual quantum mechanics formalism in configuration space. Finally, we delve into the Aharonov-Bohm effect, employing a density operator to depict states with positional and momentum uncertainty, showcasing its manifestation through distinctive interference patterns in the temporal evolution of Wigner functions under an electric potential, and emphasizing the intrinsically quantum nature of this phenomenon.}




\maketitle

\clearpage
\section{Introduction}\label{sec0}
The Aharonov-Bohm effect is the name given to a phenomenon in which the wave function of a charged particle experience a phase variation due to the influence of electromagnetic potentials, in an area with null electromagnetic fields. In 1949, Werner Ehrenberg and Raymond Siday described for the first time the Aharonov-Bohm effect \cite{I.E-S}; but it was not until 1959, when the effect was proved by David Bohm and Yakir Aharonov \cite{I.A-B}, emphasizing the physical meaning of the electromagnetic potentials. Up to that moment, they were regarded as mere mathematical objects used to simplify the computation of electromagnetic systems. It was not until 1982, when this effect could be measured by using a holography electron microscope (Tonomura \textit{et al.} \cite{I.A-B3}). More recently, the Aharonov-Bohm effect has been used for different purposes, as the observation of gauge fields \cite{Exp.A-B1} or the measure of local deformations on graphene \cite{Exp.A-B2}.\\

The setup to prove the Aharonov-Bohm effect has consisted in measuring the corresponding phase variation. For such purpose, an original wave packet has been divided into two equal components and making each of them go through an area where different potentials (scalar or vector) are acting. When recombined and measured, a phase difference among both components is evidenced. 

However, this approach for understanding the Aharonov-Bohm effect is not possible for the majority of the phase-space descriptions since they are based on real quasiprobability distributions. In this work, we propose several solutions to study the effect by using genuine phase-space formalisms. Indeed, from the beginning of the XX century, quantization has played a central role in almost all physics fields. In the beginning, after the pioneer work by Heisenberg \cite{art:heisenberg} in 1927, where the principle of uncertainty was postulated as one of the cornerstones of quantum mechanics, it seemed quite unnatural to describe it by using any phase-space formalism. This description, used in classical mechanics, treats on equal footing position and momentum. The introduction of phase space as a valid quantum formalism was first established by Wigner \cite{Wigner} in 1932, when he proposed a quasiprobability distribution function in the derivation of quantum correction terms for the Boltzmann equation. This function would be later known as the Wigner function, key for this new way of describing quantum phenomena. The same year, Weyl \cite{art:weyl} developed the correlation between phase-space functions and \textit{Weyl ordered} operators, in such a way that the Wigner functions can be described as the Weyl transform density matrices of quantum systems. Nevertheless, it was not until 1946 when the theory was considered consistent, owing to the contributions made by Groenewold and Moyal.\\

Groenewold \cite{GROENEWOLD1946405} published an article in which he proved that the Weyl correspondence was in fact an invertible transformation instead of a quantization rule. Thus, the $\star-\text{product}$ was defined as basis for this isomorphism. At the same time, Moyal \cite{moyal_1949} established the same theory by studying expected values of monomials with the structure $q^n p^m$. In 1949, the theory was finally regarded as complete. During the second half of the XX century, authors like Takabayasi (1954) \cite{Takabayasi}, Baker (1958) \cite{PhysRev.109.2198}, Fairlie (1964) \cite{fairlie_1964}, and Kubo (1964) \cite{19642127} have found important applications for this formalism as well as developing more logic aspects of it. \\

In this work, we characterize for the first time the Aharonov-Bohm effect within two different phase-space formalisms such as the one associated to the quasiprobability Wigner function and the one defined within the Segal-Bargmann space. For such purpose, the text will be divided into two different sections. The first of them will briefly summarize these two quantum mechanics formalisms; that is, it will define mathematical tools related with expected values and time evolution of the systems. The second part will be devoted to the description of several particular cases in which Aharonov-Bohm effect is present and can be described by using the previously discussed descriptions. 
\clearpage
\section{Quantum mechanics phase-space formulation}\label{sec1}

The formulation of quantum mechanics in  phase space is mainly based in the use of distribution functions. In part, given that the common probability of position and momentum is not compatible with the uncertainty principle, this functions must only be regarded as mathematical tools. The most famous distribution function is the aforementioned Wigner function. Nevertheless, there exist other examples, such as the Glauber-Sudarshan (1963) \cite{PhysRevLett.10.277, PhysRev.131.2766}  or Husimi (1940) \cite{I.husimi} functions. The advantage of using these distributions over other formalisms lies in the fact that they only involve $c$-number equations and not operators. Such distribution functions in phase space $\left[F^f(q,p)\right]$, can be defined directly through the density operator $\hat{\rho}$, depending on the position $\hat{Q}$, and momentum $\hat{P}$, operators and time $t$, as \cite{qps-distrib}
\begin{eqnarray}
    &&\operatorname{Tr}\left\{\hat{\rho}(\hat{Q}, \hat{P}, t) \text{\large e}^{i \xi \hat{Q}+i \eta \hat{P}} f(\xi, \eta)\right\}\nonumber \\
&&\;\;\;\;\;\;
=\int d q \int d p \ \text{\large e}^{i \xi q+i \eta p} F^{f}(q, p, t)\,,
\end{eqnarray}
such that
\begin{eqnarray}
&F^f(q,p)=& \int d \xi  d \eta  d q^{\prime}\left\langle q^{\prime}+\frac{\eta \hbar}{2} \right \rvert \hat{\rho} \left \rvert q^{\prime}-\frac{\eta \hbar}{2} \right\rangle \nonumber \\ &&  \cdot\frac{f(\xi, \eta)}{4 \pi^{2}} \text{\large e}^{i[ \xi\left(q^{\prime}-q\right)-\eta p]}\,,
\end{eqnarray}
where $f(\xi, \eta)$ is related to the different association rules of operators describing the system in Hilbert space ($\mathcal{H}$). In Table \ref{tab:distribfunction}, we summarize the rules of association of the distribution functions that we will discuss along this work.

\begin{table*}[t]
\centering
\scalebox{0.85}{%
\begin{tabular}{lll}
\hline
\textbf{Distribution functions} & \textbf{Rule of association} &  $\mathbf{f }$ \\ \hline \vspace{-2.5mm}\\
 Wigner ($W$)& \begin{tabular}[c]{@{}l@{}}Weyl \\ \\ $\left(e^{i \xi q+i \eta p} \leftrightarrow e^{i \xi \hat{Q}+i \eta \hat{P}}=e^{z \hat{a}^{\dagger}-z^{*} \hat{a}}\right)$\end{tabular} &  1 \vspace{1mm}\\ \hline \vspace{-2.5mm}\\
Husimi ($H$) & \begin{tabular}[c]{@{}l@{}}generalized antinormal\\ \\ $\left(e^{i \xi q+i \eta p}=e^{v \beta^{*}-v^{*} \beta}\right. \left.\leftrightarrow e^{-v^{*} b} e^{v b^{\dagger}}\right)$\end{tabular} & $e^{-\hbar\left( \xi^{2} / m \kappa+ m \kappa \eta^{2}\right)/ 4} =e^{-\rvert v\rvert^{2} / 2}$ \vspace{1mm}\\ \hline
\end{tabular}}
\caption{Types of distribution functions according to the association rule of operators defined in the Hilbert space $\mathcal{H}$  \cite{qps-distrib}.
}
\label{tab:distribfunction}
\end{table*}

\subsection{Phase-space quantization}\label{subsec2.1}

In the realm of quantum mechanics, we focus on two pivotal observables: position ($\hat{Q}$) and momentum ($\hat{P}$). These observables obey the canonical commutation relations, notably $[\hat{Q},\hat{P}]=i\hbar$. To work with them effectively, we express them in both position and momentum representations.

However, this poses an initial challenge. How can we reconcile these seemingly incompatible representations in a unified phase-space formalism? The answer lies in the Weyl quantization method, as proposed by Weyl. This method provides the transformation of phase-space functions into operators in Hilbert space through the Weyl transform. This transform is defined as \cite{art:weyl}:
\begin{eqnarray}
    &\hat{A}(\hat{Q},\hat{P})&=T_W[A]\nonumber\\
    &&=\int dq dp d\xi d\eta \frac{A(q,p)}{(2\pi\hbar)^2} \text{\large e}^{\frac{i}{\hbar}[\xi(\hat{Q}-q)+\eta(\hat{P}-p)]},\nonumber\\
    &&
\end{eqnarray}
where $A$ is any given function of the phase space, and $\hat{A}$ is its corresponding operator in Hilbert space (expressed in terms of position and momentum operators).

Moreover, the Weyl transform is reversible. We can, therefore, obtain alternative expressions for phase-space functions using the inverse Weyl transform:

 \begin{equation}\label{eq_TFG:TW_pos}
    A(q,p)=T_W^{-1}[\hat{A}]=\int dy \ \text{\large e}^{-i\frac{py}{\hbar}} \braket{q+\frac{y}{2}}{\hat{A}}{q-\frac{y}{2}}.
\end{equation}

This duality allows us to effortlessly switch between position and momentum representations.

To preserve the canonical commutation relations in phase space, we introduce the Moyal product. This product, derived from the inverse Weyl transform, not only ensures the consistency of commutation relations but also has a concise representation using "\textit{Bopp shifts}":

\begin{eqnarray}
    &A(q,p)&\star B(q,p)=\nonumber\\
    &&=A\left(q+\frac{i\hbar}{2}\partial_p,p-\frac{i\hbar}{2}\partial_q\right) B\left(q,p\right)\hspace{1cm}\nonumber\\ \nonumber\\
    &&=B\left(q-\frac{i\hbar}{2}\partial_p,p+\frac{i\hbar}{2}\partial_q\right) A\left(q,p\right).
\end{eqnarray}

The Moyal bracket \cite{moyal_1949}, defined as $[A,B]_M \equiv A\star B - B\star A$, preserves the commutation relations in phase space:

\begin{equation}
[q,p]_M = q\star p - p\star q = (qp+i\hbar/2) - (pq-i\hbar/2) = i\hbar.
\end{equation}

This proves that this quantum mechanics formalism satisfies the canonical commutation relations.

Now, let us delve into the Wigner function, a fundamental concept in this formalism. The Wigner function is derived from the density operator and plays a crucial role in characterizing quantum states in phase space. It offers a bridge between the quantum world and classical phase space. Notably, it exists in both position and momentum representations, providing unique insights into a quantum properties of the system.

The Wigner function \cite{Wigner, Cutright} is defined as :
\begin{equation}\label{eq_TFG:Wigner_pos}
W(q,p) = \frac{1}{2\pi\hbar}\int dy\, \text{\large e}^{-\frac{i}{\hbar}py}\psi\left(q+\frac{y}{2}\right)\psi\left(q-\frac{y}{2}\right)^* ,
\end{equation}
where $\psi(q)$ is the wave function of the system in position representation, defined as $\psi(q) = \bra{q}{\psi}\rangle$. We can also express the Wigner function in terms of the wave function in momentum representation \cite{Cutright}:

\begin{equation}\label{eq_TFG:Wigner_mom}
    W(q,p)=\frac{1}{2\pi\hbar}\int du\ \text{\large e}^{i\frac{qu}{\hbar}}\Tilde{\psi}\left(p+\frac{u}{2}\right)\Tilde{\psi}\left(p-\frac{u}{2}\right)^*.
\end{equation}

The Wigner function is a quasiprobability distribution, enabling us to explore the probabilistic nature of quantum systems. It can take negative values, a characteristic that arises due to the uncertainty principle, preventing us from assigning a definite physical meaning to all points in phase space.
 
The time-dependent Schrödinger equation transforms into an equation governing the evolution of the Wigner function via the Moyal bracket. This allows us to track the dynamic behaviour of quantum systems in phase space, including special cases involving stationary states,
\begin{equation}\label{eq_TFG:ev_wigner}
    \frac{\partial W}{\partial t}=\frac{1}{i\hbar}[H,W]_M\,.
\end{equation}

In cases where density operators are not diagonal, the Wigner function still holds significance. It can be constructed from a superposition of Wigner functions associated with individual pure states, offering insights into the energy eigenstates and their respective energy values \cite{L.},
\begin{eqnarray}\label{eq_TFG:no_diag_wigner}
    H\star W_{i,j}&=E_i\ W_{i,j},\nonumber\\
    W_{i,j}\star H&=E_j\ W_{i,j}.
\end{eqnarray}

In summary, the phase-space formalism in quantum mechanics, anchored by observables, the Weyl quantization method, Moyal product, and the versatile Wigner function, provides a comprehensive framework for understanding quantum systems and their evolution in phase space.

  \subsection{Segal-Bargmann space formalism of quantum mechanics}\label{sec3}

Segal-Bargmann space \cite{I.Bargmann, I.Segal, Y.1.9}, denoted as $\mathcal{H} L^{2}(U, \alpha)$, is a mathematical space consisting of square integrable, holomorphic functions over an open set $U \subset \mathbb{C}$ with respect to the weight $\alpha(z)=e^{-\rvert z\rvert^2}/\pi$.

In $\mathcal{H} L^{2}(U, \alpha)$, an inner product can be defined, represented as:

\begin{equation}
\langle f \mid g\rangle=\int_{U} d z\ \alpha(z) f(z)^{*} g(z), \quad f, g \in \mathcal{H} L^{2}(U, \alpha).
\end{equation}

This inner product forms the basis for various connections to quantum mechanics. Indeed, it can be proved that $\left\{z^{n}\right\}_{n=0}^{\infty}$ is a valid basis of Segal-Bargmann space, given that holomorphic functions are always analytic and hence they can be expressed as a power series  \cite{Y.1.T.}.

In the context of quantum mechanics, creation and annihilation operators in Hilbert space are derived from the dimensionless position and momentum operators as:

\begin{eqnarray}\label{eq:20}
&\hat{a}=\dfrac{1}{\sqrt{2}}\left(\hat{Q}^{\prime}+i \hat{P}^{\prime}\right), \nonumber\\
&\hat{a}^{\dagger}=\dfrac{1}{\sqrt{2}}\left(\hat{Q}^{\prime}-i \hat{P}^{\prime}\right).
\end{eqnarray}

The commutation relations for these operators are analogous to those of $z$ and $\partial_z$ in Segal-Bargmann space. This connection is established through the Segal-Bargmann transform \cite{SvN2, SvN}.

In Segal-Bargmann space, the Husimi function \cite{I.husimi} $\mathscr{H}(z)$ is a quasiprobability distribution defined as:

\begin{equation}
\mathscr{H}(z)=\frac{\text{\large e}^{-\left \rvert z\right \rvert^{2}}}{2 \pi}\left \rvert \phi(z)\right \rvert^{2}.
\end{equation}

The Husimi function possesses several noteworthy properties: it is real, non-negative, bounded, has no marginal distributions, and is normalized.

When considering the time evolution of quantum systems in Segal-Bargmann space, creation and annihilation operators play a crucial role. The Hamiltonian operator $\hat{H}$ is constructed using these operators. The time-dependent Schrödinger equation in this space is given by:

\begin{equation}\label{eq_TFG:ev_temp_segal}
\frac{\partial \phi}{\partial t}=\frac{1}{i \hbar} T_{S B}[\hat{H}] \phi .
\end{equation}

Energy eigenstates in Segal-Bargmann space evolve over time with a global phase given by:

\begin{equation}
\phi(z, t)=e^{-i E t / \hbar} \phi(z, 0).
\end{equation}

These fundamental definitions and equations establish the significance of Segal-Bargmann space in quantum mechanics, providing a bridge between holomorphic functions and quantum states.

\section{Aharonov-Bohm effect}\label{sec4}

Foremost, we would like to clarify the units we will use for discussing the Aharonov-Bohm effect. On the one hand, we will use natural units determined by $\hbar=c=\epsilon_0=1$. On the other hand, although we reviewed the Segal-Bargmann space by using dimensionless operators in the previous section, we will introduce dimensional operators in the following sections. The question is that we aim to obtain measurable results corresponding to the Aharonov-Bohm effect. Therefore, we will be using the constant $m$ (the proposed particle mass for all the following study cases) in order to endow the proper dimensions to the corresponding operators. Consequently, we will take any operator with dimensions $\hat{A}$ and obtain its representation in Segal-Bargmann space as \begin{equation}
T_{S B}[\hat{A}]=m^{n} T_{S B}\left[\hat{A}^{\prime}\right],
\end{equation} where $\hat{A}'$ is the dimensionless operator and $n$ is a number such that $m^n$ gives the proper dimension to the dimensional $\hat{A}$ operator.\\

Before approaching Aharonov-Bohm effect in the different proposed formalisms, it is convenient to solve the case for a free particle system\footnote{It must be emphasized the fact that plane waves are a pathological case when
normalized. However, we will use them since they allow to simplify the discussion of the Aharonov-Bohm effect.}.\\

\subsection{Free particle}\label{subsec4.1}
The Hamiltonian operator for a massive particle in the absence of interactions within the Segal-Bargmann space is given by 
\begin{equation}
\hat{H}=\frac{\hat{P}^{2}}{2 m}.\label{hamiltonianfree}
\end{equation}

Momentum operator in the Segal-Bargmann space has the form $T_{S B}[\hat{P}]=i m\left(z-\partial_{z}\right) / \sqrt{2}$, while position operator takes the form $T_{S B}[\hat{Q}]=\dfrac{1}{\sqrt{2}}\left(z+\partial_{z}\right)$. Hence, the canonical commutation relations are ensured: $[Q, P]=\left[\left(z+\partial_{z}\right) / \sqrt{2}, i\left(z-\partial_{z}\right) / \sqrt{2}\right]=i$. Furthermore, the Hamiltonian operator is expressed as
\begin{eqnarray}
&T_{S B}[\hat{H}]&=-\frac{m}{4}\left[\left(z-\partial_{z}\right)\left(z-\partial_{z}\right)\right]\nonumber\\ &&=-\frac{m}{4}\left[z^{2}+\partial_{z}^{2}-2 z \partial_{z}-1\right].
\end{eqnarray}

The particle state, in Segal-Bargmann space, is linked to the square-integrable holomorphic function $\phi(z)$. This function must fulfil the eigenvalue equation
\begin{eqnarray}
&T_{S B}[\hat{H}] \phi(z)&=-\frac{m}{4}\left[z^{2}+\partial_{z}^{2}-2 z \partial_{z}-1\right] \phi(z)\nonumber\\
&&=E \phi(z).
\end{eqnarray}

One solution to this eigenvalue equation is
\begin{equation}
\phi(z)=A \text{\large e}^{\frac{z}{2}(4 i \sqrt{E / m}+z)}+B \text{\large e}^{\frac{z}{2}(-4 i \sqrt{E / m}+z)},
\end{equation}
where $A$ and $B$ are constants. Since the Hamiltonian operator and the momentum operator commute ($[\hat{H},\hat{P}]=0$), it must exist an eigenfunction basis common to both operators.\\
Applying the momentum operator to the previous expression, we obtain
\begin{eqnarray}
&T_{S B}[\hat{P}] &\cdot A \text{\large e}^{\frac{z}{2}(4 i \sqrt{E / m}+z)}\nonumber \\ &&=\sqrt{2 E m} \cdot A \text{\large e}^{\frac{z}{2}(4 i \sqrt{E / m}+z)},\hspace{1cm} \nonumber\\
&T_{S B}[\hat{P}] &\cdot B \text{\large e}^{\frac{z}{2}(-4 i \sqrt{E / m}+z)}\nonumber \\&&=-\sqrt{2 E m} \cdot B \text{\large e}^{\frac{z}{2}(-4 i \sqrt{E / m}+z)}.
\end{eqnarray}

In conclusion, the eigenfunctions are associated with the different senses of the direction of the momentum:
\begin{eqnarray}
\phi_{+}(z)&=&A \text{\large e}^{\frac{z}{2}(4 i \sqrt{E / m}+z)}, \nonumber\\
&\text{with}&p=\sqrt{2 m E}\,; \nonumber\\ \phi_{-}(z)&=&B \text{\large e}^{\frac{z}{2}(-4 i \sqrt{E / m}+z)},\nonumber\\ &\text{with}&p=-\sqrt{2 m E}\,.
\end{eqnarray}

On the other hand, we can study the same system by using the Wigner function. The Hamiltonian operator of a massive particle in absence of interaction is given by Equation \eqref{hamiltonianfree}. The momentum operator in phase space is given by $T_{w}^{-1}[\hat{P}]=p$. Consequently, the canonical commutation relations (with the commutators belonging to this space, Moyal brackets) are fulfilled  $[\hat{Q}, \hat{P}]=[q, p]_{M}=q \star p-p \star q=i$. 
Therefore, the Hamiltonian operator can be written as 
\begin{eqnarray}
    &T_{w}^{-1}[\hat{H}] \star=\frac{p^{2}}{2 m} \star&=\dfrac{1}{2 m}\left[\left(p-\dfrac{i}{2} \partial_{q}\right)\left(p-\dfrac{i}{2} \partial_{q}\right)\right]\nonumber\\
    &&=\dfrac{1}{2 m}\left[p^{2}-\dfrac{1}{4} \partial_{q}^{2}-i p \partial_{q}\right].
\end{eqnarray}
The particle state, in phase space, is linked to the Wigner function, $W(q, p)$. This function must satisfy the eigenvalue equation,
\begin{eqnarray}
&T_{w}^{-1}[\hat{H}] &\star W(q, p)\nonumber\\
&&=\frac{1}{2 m}\left[p^{2}-\frac{1}{4} \partial_{q}^{2}-i p \partial_{q}\right] W(q, p)\nonumber \\ &&=E\, W(q, p).
\end{eqnarray}
Owing to the fact that Wigner functions must be real by definition, by taking the conjugate of the evolution equation, we can write
\begin{eqnarray}
&&{\left[\left(p^{2}-2 m E\right)-\dfrac{1}{4} \partial_{q}^{2}-i p \partial_{q}\right] W(q, p)=0}, \nonumber\\
&&{\left[\left(p^{2}-2 m E\right)-\dfrac{1}{4} \partial_{q}^{2}+i p \partial_{q}\right] W(q, p)=0 .}\hspace{8mm}
\end{eqnarray}
By subtracting both expressions, we obtain
\begin{equation}
2 i p\, \partial_{q} W(q, p)=0.
\end{equation}
Therefore, if $p\neq 0$, the Wigner function does not depend on position. Substituting again in the evolution equation:
\begin{equation}
\left(p^{2}-2 m E\right) W(p)=0\,. 
\end{equation}
It means that we can write two different solutions by distinguishing the sense of the momentum: 
\begin{equation}
W(p)=A_w \delta(p-\sqrt{2 m E})+B_w \delta(p+\sqrt{2 m E}),
\end{equation}
where $A_w$ and $B_w$ are constants. As we discussed for the Segal-Bargmann space, since the Hamiltonian operator and momentum operator commute ($[\hat{H},\hat{P}]=0$),  it must exist an eigenfunction basis common to both operators. Contrary to what we did in Segal-Bargmann space, applying the momentum operator is not helpful, since its eigenvalue is a phase-space variable. However, the only values of such variable for which the Wigner function is not null are $p=\pm \sqrt{2 m E}$. Therefore, we can conclude that
\begin{eqnarray}
W_{+}(q, p)&=&A_w \delta(p-\sqrt{2 m E}), \nonumber\\
&&\text{what means}\;\;\;p=\sqrt{2 m E}; \nonumber\\ W_{-}(q, p)&=&B_w \delta(p+\sqrt{2 m E}), \nonumber\\
&&\text{what means}\;\;\;p=-\sqrt{2 m E} .
\end{eqnarray}

\subsection{Aharonov-Bohm effect with non-zero electric potential}\label{subsec4.2}
We will assume that the system is formed by two regions in which it is possible to have constant electric potentials $\varphi_i$ that differ between them. We send a charged particle, with charge $q$ and energy $E_0$, whose probability function will propagate through both regions. At a time $t=0$, the electric system will be connected and at $t=\tau>0$, it will be disconnected \footnote{It is supposed that the activation and disconnection processes will not affect the particle evolution.}. In such a case, the Hamiltonian operators corresponding to both regions are \cite{I.J-S}
\begin{eqnarray}
&\hat{H}_{1}=\dfrac{\hat{P}^{2}}{2 m}+q \varphi_{1}, \nonumber\\
&\hat{H}_{2}=\dfrac{\hat{P}^{2}}{2 m}+q \varphi_{2} .
\end{eqnarray}

   In Segal-Bargmann space, we denote $\phi_1(z,t)$ as the quantum state that evolves with $T_{S B}[\hat{H}_{1}]$, and $\phi_2(z,t)$ as the one that evolves with $T_{S B}[\hat{H}_{2}]$. Therefore, the total state is described by $\phi(z, t)=\left(\phi_{1}(z,t)+\phi_{2}(z,t)\right) / 2$. We begin with a plane-wave, so the initial condition is $\phi(z,0)=\phi_{1}(z,0)=\phi_{2}(z,0)=\phi_{+}(z)$ (for simplicity, we are taking the positive momentum solution, although it is possible to do the same analysis with the negative momentum solution). Once the electrical system is connected, $0<t<\uptau$, each component $\phi_i(z,t)$ acquires a different energy given by
\begin{eqnarray}
&T_{S B}\left[\hat{H}_{1}\right] \phi_{1}(z,t)=E_{1} \phi_{1}(z,t), & E_{1}=E_{0}+q \varphi_{1}, \nonumber\\
&T_{S B}\left[\hat{H}_{2}\right] \phi_{2}(z,t)=E_{2} \phi_{2}(z,t), & E_{2}=E_{0}+q \varphi_{2}.\nonumber\\
&&
\end{eqnarray}

From the expression we obtained for the time evolution of systems in Segal-Bargmann space \eqref{eq_TFG:ev_temp_segal}, we can write
 \begin{eqnarray}
&\phi(z, t)&=\frac{1}{2}\left[\text{\large e}^{-i E_{1} t} \phi_{1}(z,0)+\text{\large e}^{-i E_{2} t} \phi_{2}(z,0)\right]\nonumber \\&&=\frac{\text{\large e}^{-i E_{2} t}}{2} \phi_{+}(z)\left[1+\text{\large e}^{i \Delta E\cdot t}\right],
\end{eqnarray}
where $\Delta E=E_{2}-E_{1}=q\left(\varphi_{2}-\varphi_{1}\right)=q \Delta \varphi$.\\

    For a time $t>\tau$, the total probability of detecting the particle with respect to itself at $t=0$ is given by

\begin{eqnarray}
&\mathcal{P}(t) &=\frac{\rvert\phi(z, \tau)\rvert^{2}}{\rvert\phi(z, 0)\rvert^{2}}\nonumber\\
&&=\frac{1}{\left\rvert\phi_{+}(z)\right\rvert^{2}} \int d z\ \text{\large e}^{-\rvert z\rvert^{2}}\rvert\phi(z, \tau)\rvert^{2} \nonumber\\
&&=\frac{1+\mathcal{R} e\left(\text{\large e}^{i q \Delta \varphi \cdot t}\right)}{2\left\rvert\phi_{+}(z)\right\rvert^{2}} \int d z\ \text{\large e}^{-\rvert z\rvert^{2}}\left\rvert\phi_{+}(z)\right\rvert^{2} \nonumber\\
&&=\frac{1}{2}[1+\cos (q \Delta \varphi \cdot \tau)].
\end{eqnarray}

Note that we propose an open system (we allow a sudden appearance and disappearance of different electrical potentials in each conduit) so that there is no conservation of total probability in the system. Hence, we can take the change in the total probability, $\mathcal{P}(t)$, as the observable that shows the Aharonov-Bohm effect.

Provided that $q \Delta \varphi \cdot \tau \neq 2 \pi n$ with $n \in \mathbb{Z}$, the presence of the electric potential will have a measurable effect even when the electric field is zero $\left(E=-\partial_{x} \varphi=0\right)$ in the region accessible to the particle. It must be noted, that the predicted phase difference in Schrödinger formalism, for a non-zero electric potential, is $-q \int d t \varphi$. Particularizing this result to our system ($\varphi$ constant), it becomes $-q\varphi\tau$. Because of this, the phase difference between both wave functions would be $q \Delta \varphi \cdot \tau$, which is the same as the one we have obtained.

On the other hand, if the quantum state that described our system is $\ket{\psi}=(\ket{\psi_1}+\ket{\psi_2})/2$, where $\ket{\psi_i}$ is the quantum state that evolves with $\hat{H}_i$; then the density operator of our system is $\hat{\rho}=\ket{\psi}\bra{\psi}=(\hat{\rho}_{11}+\hat{\rho}_{12}+\hat{\rho}_{21}+\hat{\rho}_{22})/4$, with $\hat{\rho}_{ij}=\ket{\psi_i}\bra{\psi_j}$. Therefore, the Wigner function that described our system in the phase-space formalism is $W(x,p,t)=T_W^{-1}[\hat{\rho}]=(W_{11}(x,p,t)+W_{12}(x,p,t)+W_{21}(x,p,t)+W_{22}(x,p,t))/4$, with $W_{i,j}(x,p,t)=T_W^{-1}[\hat{\rho}_{i,j}]$. We begin again with a plane-wave, so the initial condition is $W(x,p,0)=W_{i,j}(x,p,0)=W_+(x,p)$, since we assume the positive momentum solution. Recalling Equations \eqref{eq_TFG:ev_wigner} and \eqref{eq_TFG:no_diag_wigner}, we obtain the Wigner function for the system at a certain time, $0<t<\uptau$, as
\begin{equation}
W(x, p, t)=\frac{W_+(x, p)}{2}[1+\cos (\Delta E \cdot t)].
\end{equation}

    Therefore, for a time $t>\tau$, the total probability of detecting the particle with respect to itself at $t=0$ is given by

\begin{eqnarray}
&\mathcal{P}(t) &=\frac{\int d x d p W(x, p, \tau)}{\int d x d p W(x, p, 0)}\nonumber\\
&&=\frac{1}{2}[1+\cos (q \Delta \varphi \cdot \tau)] .
\end{eqnarray}
We derive the same result that we obtained by working within the Segal-Bargmann space; which at the same time coincides with the standard result in Schrödinger formalism.
With the aim to visualize the Aharonov-Bohm effect with non-zero electric potential more easily; we build a graphical representation making use of the results we obtained by the phase-space and the Segal-Bargmann space formalisms.

\begin{figure*}[ht]
\begin{minipage}[t]{\textwidth}
    \hspace{2cm}\begin{minipage}[t]{9.1cm}
        \begin{minipage}[t]{4cm}
            \begin{figure}[H]
                \centering               \boxed{\includegraphics[width=\columnwidth]{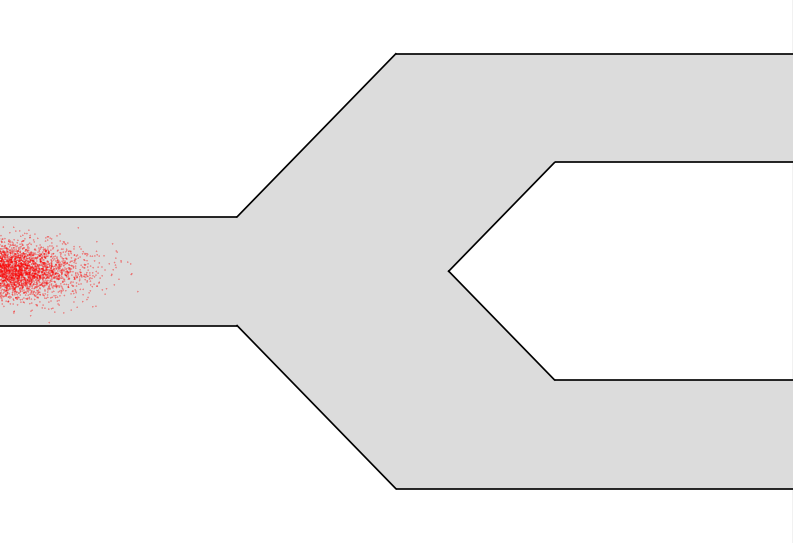}}
                
                \vspace{2mm}(a)
            \end{figure}
        \end{minipage}
        \hfill
        \begin{minipage}[t]{4cm}
            \begin{figure}[H]
                \centering                \boxed{\includegraphics[width=\columnwidth]{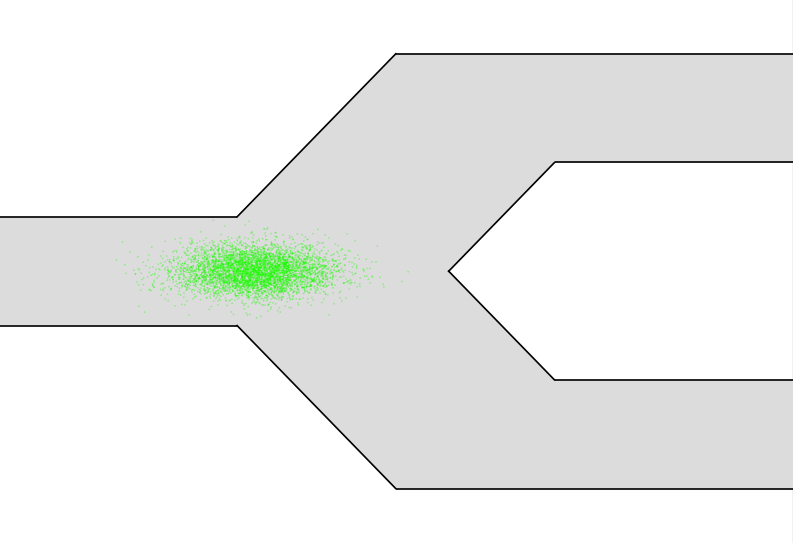}}
                
                \vspace{2mm}(b)
            \end{figure}
        \end{minipage}
        \begin{minipage}[t]{4cm}
            \begin{figure}[H]
                \centering                \boxed{\includegraphics[width=\columnwidth]{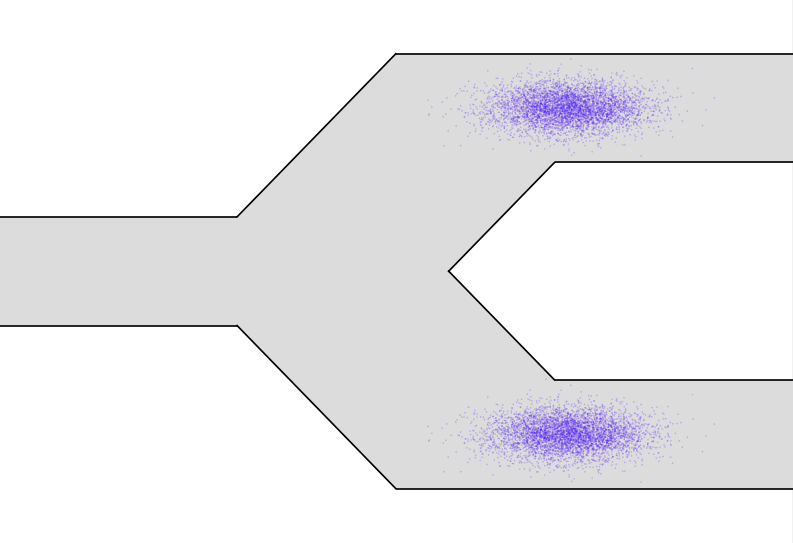}}
                
                \vspace{2mm}(c)
            \end{figure}
        \end{minipage}
        \hfill
        \begin{minipage}[t]{4cm}
            \begin{figure}[H]
                \centering                \boxed{\includegraphics[width=\columnwidth]{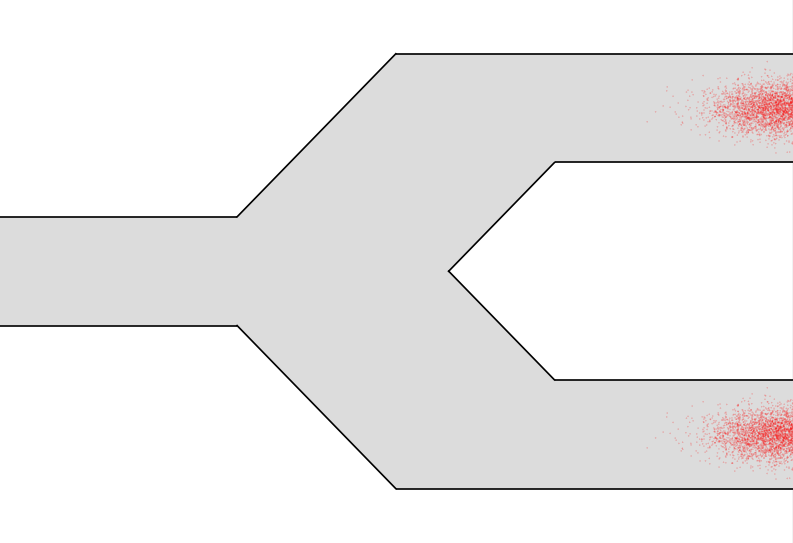}}
                
                \vspace{2mm}(d)
            \end{figure}
        \end{minipage}
    \end{minipage}
    \hfill
    \begin{minipage}[t]{2.24cm}
        \begin{figure}[H]
            \centering
            \includegraphics[width=0.8\columnwidth]{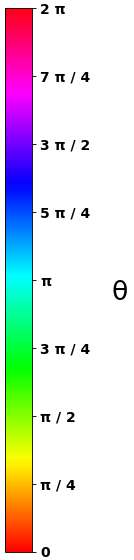}
        \end{figure}
    \end{minipage}\hspace{2cm}
    
    \vspace{2mm}
    \captionof{figure}{Graphical representation of a wave packet which evolves over time while travelling through a conduit which is forked (with zero electric potential). Although the phases correspond with a plane wave, in the graphical representation we represent the wave packet as a Gaussian localized distribution for an easier visualization. We observe in this figure that the phase of the wave packet evolves as the phase of the free particle even when it is split into two. The time interval between the frame (a) and the frame (d) is $t=\uptau=2\pi/E_0$.}\label{fig:electrico_a}
\end{minipage}
\end{figure*}

\begin{figure*}[ht]
\begin{minipage}[t]{\textwidth}
    \hspace{2cm}\begin{minipage}[t]{9.1cm}
        \begin{minipage}[t]{4cm}
            \begin{figure}[H]
                \centering               \boxed{\includegraphics[width=\columnwidth]{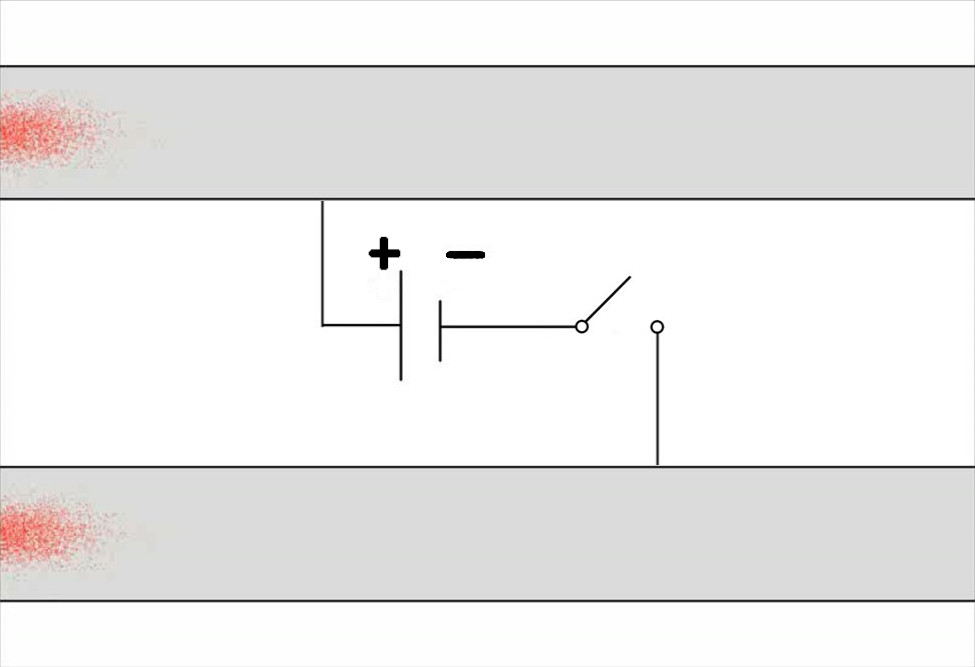}}
                
                \vspace{2mm}(a)
            \end{figure}
        \end{minipage}
        \hfill
        \begin{minipage}[t]{4cm}
            \begin{figure}[H]
                \centering                \boxed{\includegraphics[width=\columnwidth]{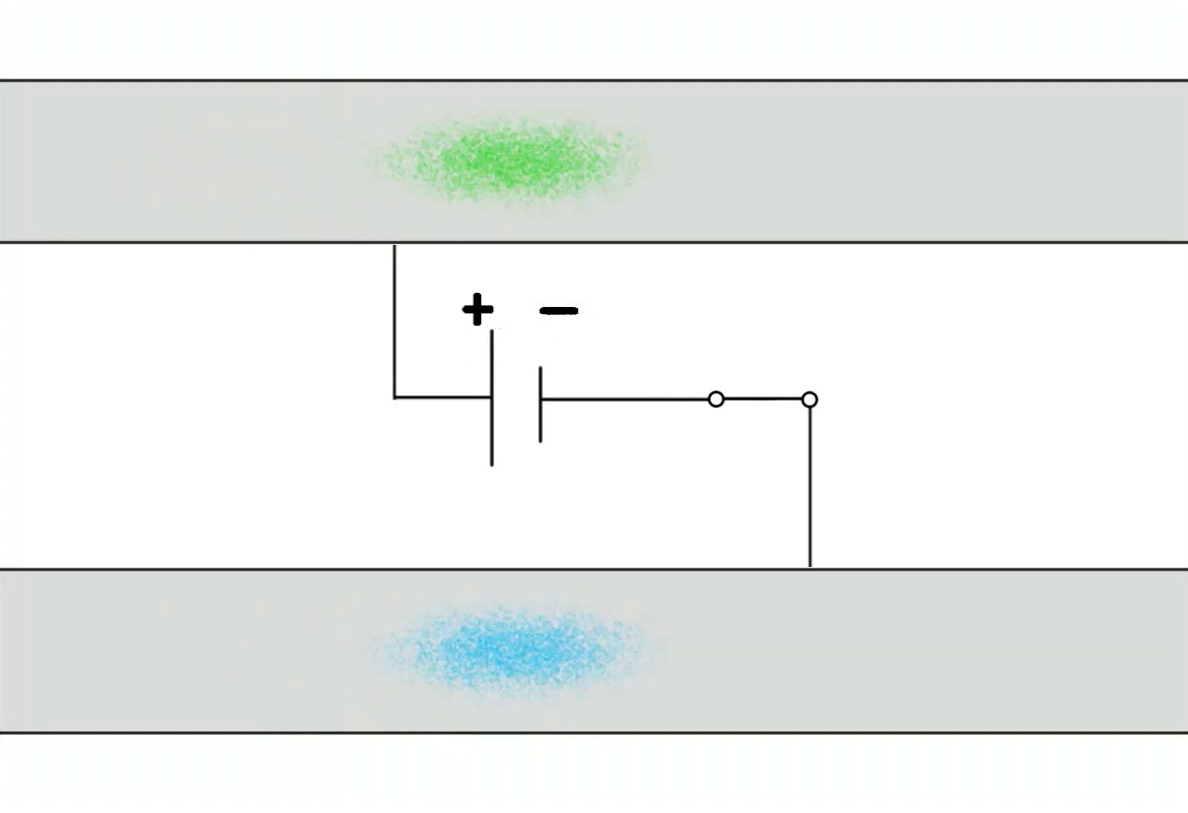}}
                
                \vspace{2mm}(b)
            \end{figure}
        \end{minipage}
        \begin{minipage}[t]{4cm}
            \begin{figure}[H]
                \centering                \boxed{\includegraphics[width=\columnwidth]{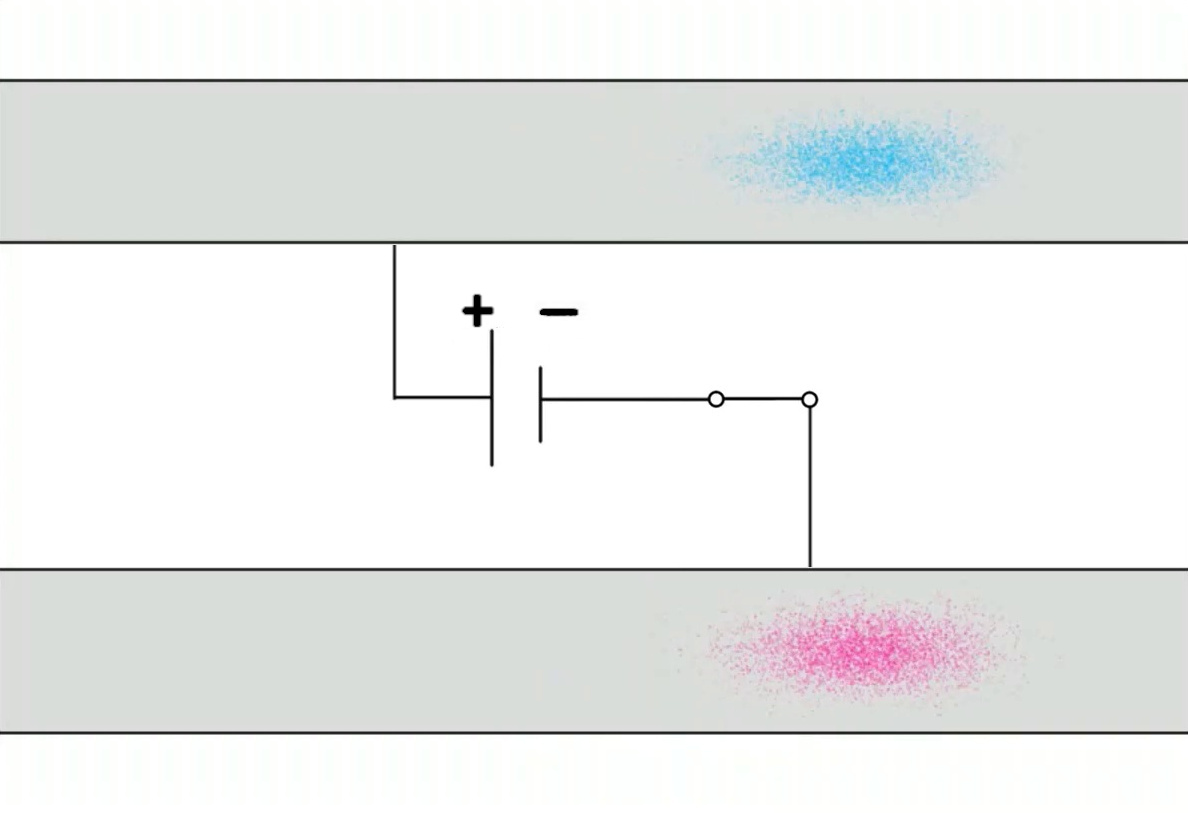}}
                
                \vspace{2mm}(c)
            \end{figure}
        \end{minipage}
        \hfill
        \begin{minipage}[t]{4cm}
            \begin{figure}[H]
                \centering                \boxed{\includegraphics[width=\columnwidth]{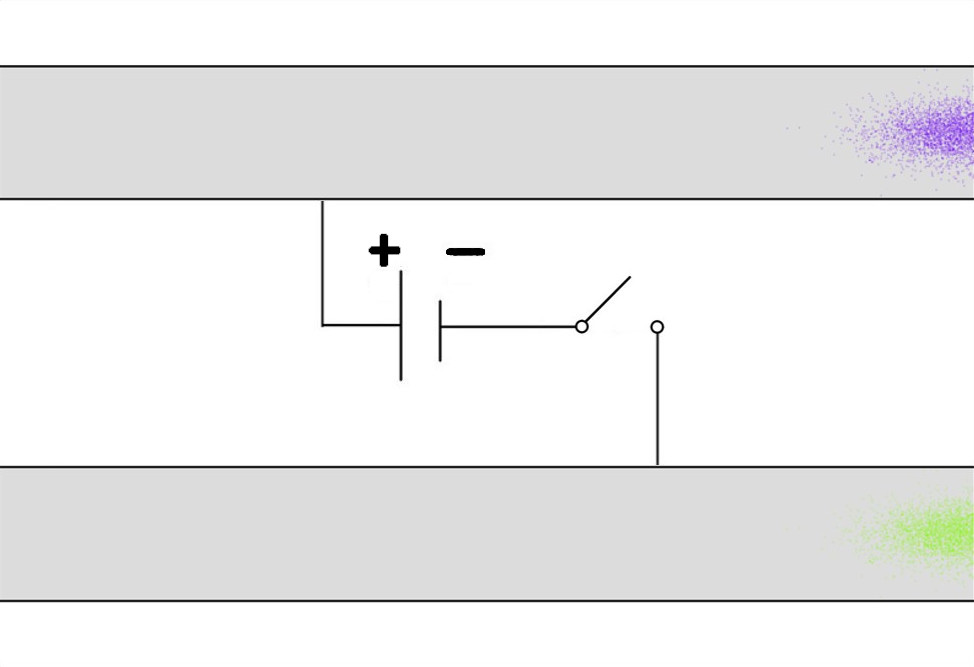}}
                
                \vspace{2mm}(d)
            \end{figure}
        \end{minipage}
    \end{minipage}
    \hfill
    \begin{minipage}[t]{2.24cm}
        \begin{figure}[H]
            \centering
            \includegraphics[width=0.8\columnwidth]{imagenes/color1.png}
        \end{figure}
    \end{minipage}\hspace{2cm}
    
    \vspace{2mm}
    \captionof{figure}{This figure starts where Fig. \ref{fig:electrico_a} finishes. Here it is represented the evolution of the wave packet through two conduits with a potential difference of $\Delta \varphi=E_0/2q$ between them. In frame (a) the electric circuit is connected, so in frame (d) (after having crossed the conduits) the phase difference between the two components is $\pi$ radians. The time interval between the frame (a) and the frame (d) is $t=\uptau=2\pi/E_0$.}\label{fig:electrico_b}
\end{minipage}
\end{figure*}

\begin{figure*}[ht]
\begin{minipage}[t]{\textwidth}
    \hspace{2cm}\begin{minipage}[t]{9.1cm}
        \begin{minipage}[t]{4cm}
            \begin{figure}[H]
                \centering               \boxed{\includegraphics[width=\columnwidth]{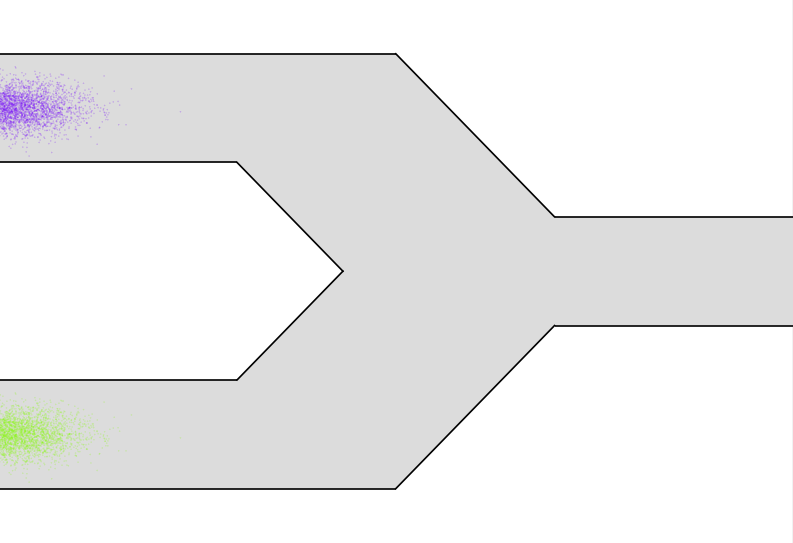}}
                
                \vspace{2mm}(a)
            \end{figure}
        \end{minipage}
        \hfill
        \begin{minipage}[t]{4cm}
            \begin{figure}[H]
                \centering                \boxed{\includegraphics[width=\columnwidth]{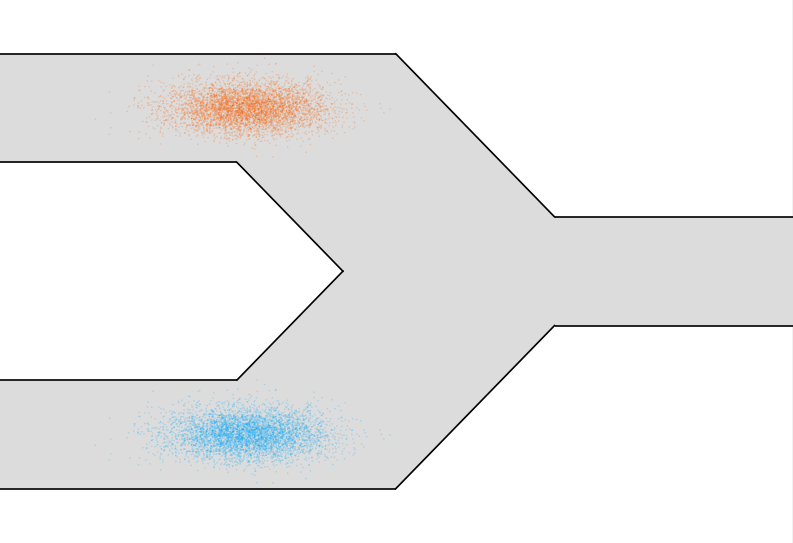}}
                
                \vspace{2mm}(b)
            \end{figure}
        \end{minipage}
        \begin{minipage}[t]{4cm}
            \begin{figure}[H]
                \centering                \boxed{\includegraphics[width=\columnwidth]{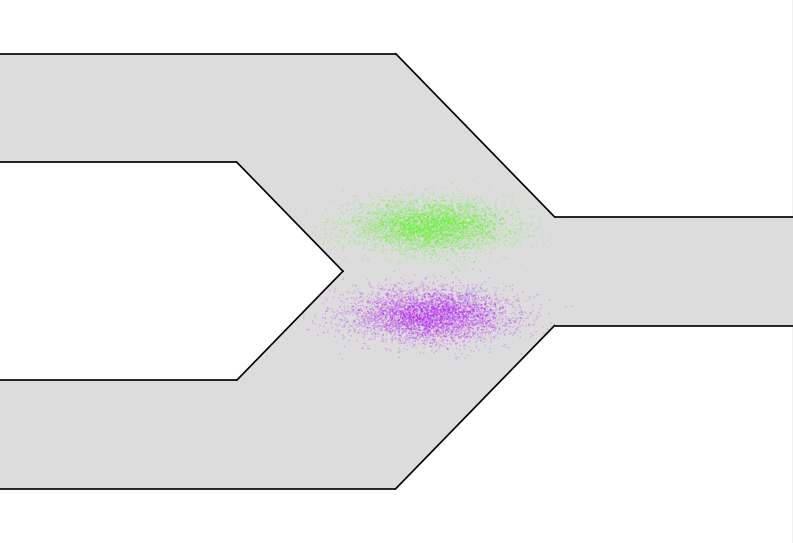}}
                
                \vspace{2mm}(c)
            \end{figure}
        \end{minipage}
        \hfill
        \begin{minipage}[t]{4cm}
            \begin{figure}[H]
                \centering                \boxed{\includegraphics[width=\columnwidth]{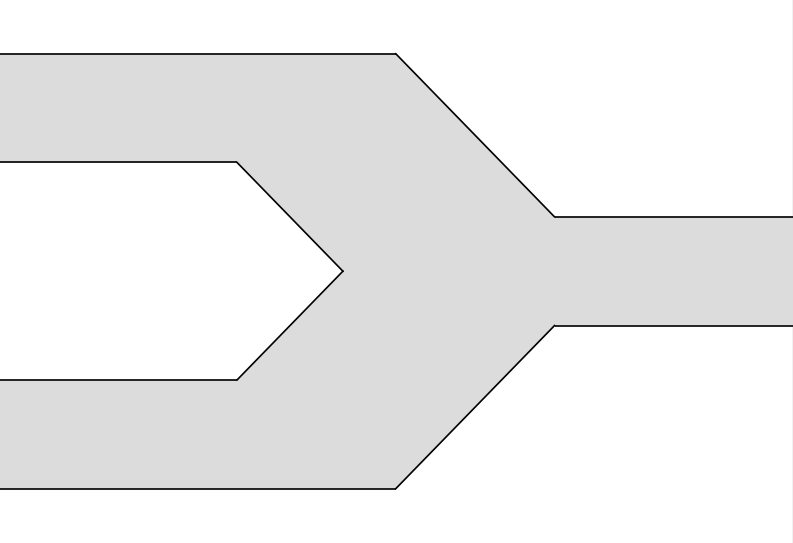}}
                
                \vspace{2mm}(d)
            \end{figure}
        \end{minipage}
    \end{minipage}
    \hfill
    \begin{minipage}[t]{2.24cm}
        \begin{figure}[H]
            \centering
            \includegraphics[width=0.8\columnwidth]{imagenes/color1.png}
        \end{figure}
    \end{minipage}\hspace{2cm}
    
    \vspace{2mm}
    \captionof{figure}{This figure starts when Fig. \ref{fig:electrico_b} finishes. Here it is represented the evolution of the wave packet through two conduits which rejoin (with null electric potential). In all frames the phase difference is $\pi$ radians since the electric potential is null. Therefore, destructive interference of the two components results when the conduits rejoin in frame (d), so the probability of finding the wave packet becomes $0$.}\label{fig:electrico_c}
\end{minipage}
\end{figure*}

In this graphical representation, both the spatial probability of finding the particle and the phase of the particle are represented through colours (the relationship between colours and the phase is presented in the sidebar). Although the results correspond with a plane wave, in the graphical representation we show the particle as a Gaussian distribution for an easier visualization. As can be observed in Fig. \ref{fig:electrico_a}, we begin with a conduit (with zero electric potential) that splits into two. The phase of the particle evolves freely; therefore, the two components are in phase at the end of the route.
Fig. \ref{fig:electrico_b} illustrates the situation for an induced potential difference between the conduits of $\Delta \varphi=E_0/2q$ during a time interval of $t=\uptau=2\pi/E_0$. Finally, in Fig. \ref{fig:electrico_c}, the conduits rejoin. The two components of the wave function superposed, and in this particular case, the probability of finding the particle becomes exactly zero.

\subsection{Aharonov-Bohm effect with non-zero magnetic vector potential}\label{subsec4.3}

In this case, we will assume an infinite solenoid with a constant magnetic field inside. Outside the solenoid, its value is zero. Therefore, the magnetic vector potential outside the solenoid can be written as
\begin{equation}
    \mathbf{A}=\frac{a^2B}{2r}\mathbf{u}_\vartheta,
\end{equation}

\noindent where $a$ is the radius of the solenoid ($a<r$) and $\vartheta$ is the angular coordinate.

Suppose a charged particle that surrounds the solenoid on its perpendicular plane at a constant radial distance of its centre, $r=R$. Just like in the previous case, the probability function will be divided into two parts; one part surrounds the solenoid clockwise, and the other one surrounds it counterclockwise.

We will change the coordinate system in order to simplify the analysis. We define a new coordinate $s\equiv R\vartheta$, so its canonical conjugate momentum is given by $p_s=m(r^2/R^2)ds/dt$. The momentum vector of the particle, expressed in the new coordinate system, is then given by $\mathbf{p}=p_r\mathbf{u}_r+p_s(R/r)\mathbf{u}_s$. Using the phase-space coordinates of our particle ($p_r=0$ y $r=R$), $s$ becomes the arc length of the circular trajectory of the particle, and $\mathbf{p}=p_s\mathbf{u}_s$ becomes its conjugate momentum. Furthermore, it is easy to verify that $\mathbf{u}_\vartheta=\mathbf{u}_s$. Hence, the magnetic vector potential and the particle momentum have both the same direction. Therefore, we can analyse the system within a two-dimensional phase space, by taking $s$ as spatial coordinate and $p$ as its canonical conjugate momentum (we omit the reference to the subscript $s$ since the momentum only has one non-zero component).

As in the electric potential case, the magnetic system is turned on at $t=0$ and is turned off at $t=\uptau>0$. While the solenoid is connected, the hamiltonian of the system is given by \cite{I.J-S}
\begin{equation}\label{eq_TFG:ham_A}
    \hat{H}=\frac{(\hat{P}-qA)^2}{2m}=\frac{1}{2m}\left[\hat{P}^2-2qA\hat{P}+q^2A^2\right].
\end{equation}

\begin{figure*}[ht]
\begin{minipage}[t]{\textwidth}
    \hspace{2cm}\begin{minipage}[t]{9.1cm}
        \begin{minipage}[t]{4cm}
            \begin{figure}[H]
                \centering               \boxed{\includegraphics[width=\columnwidth]{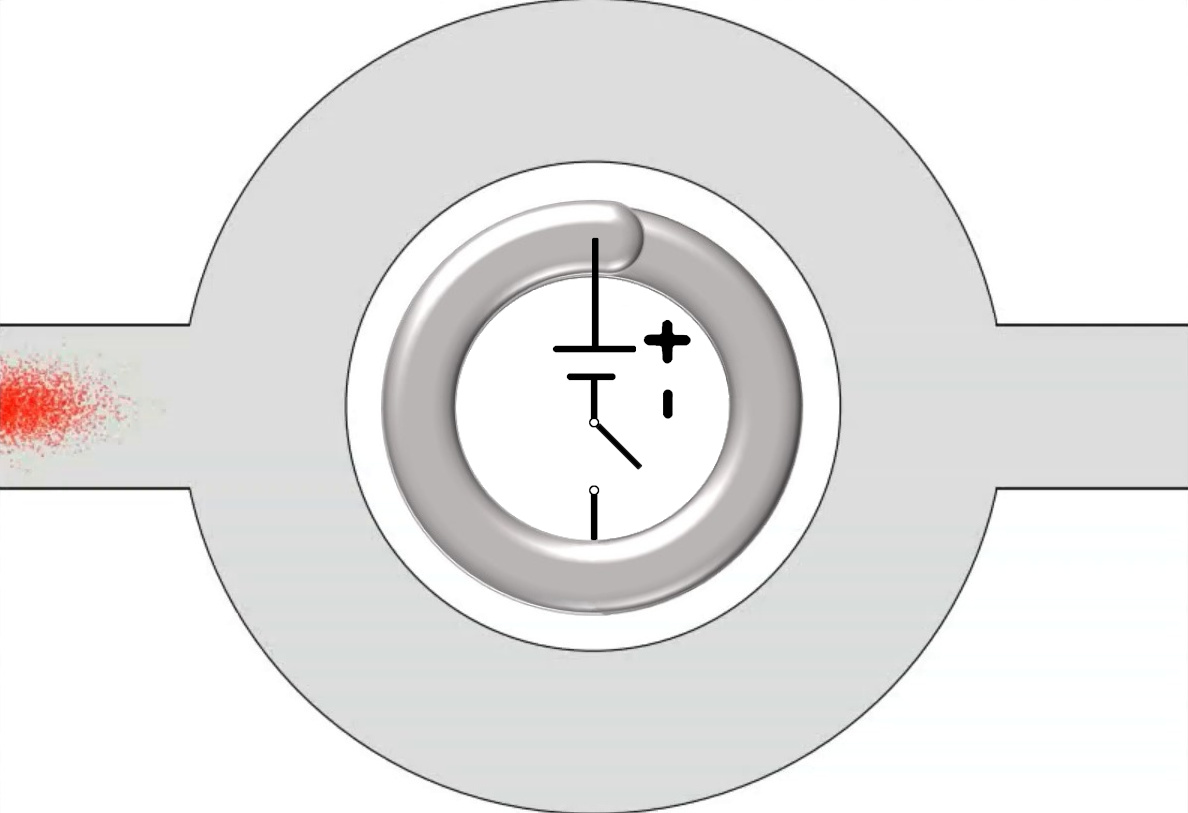}}
                
                \vspace{2mm}(a)
            \end{figure}
        \end{minipage}
        \hfill
        \begin{minipage}[t]{4cm}
            \begin{figure}[H]
                \centering                \boxed{\includegraphics[width=\columnwidth]{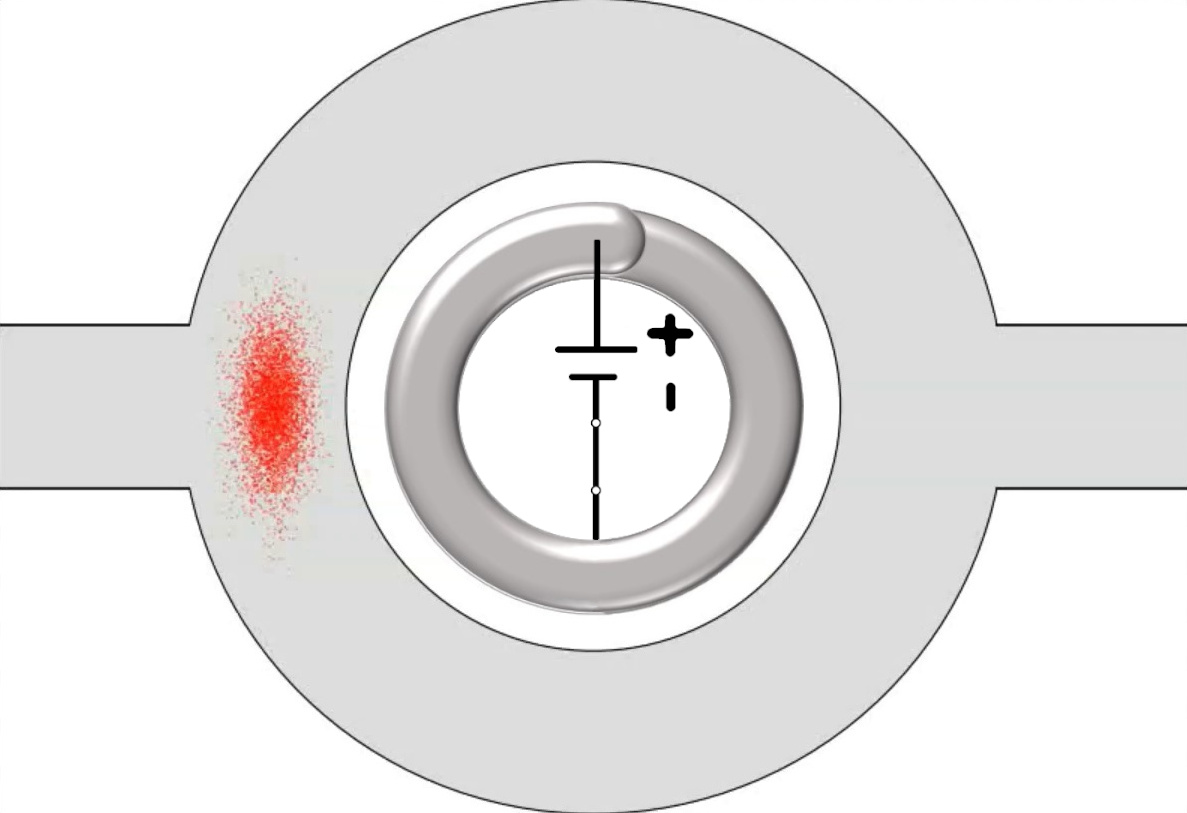}}
                
                \vspace{2mm}(b)
            \end{figure}
        \end{minipage}
        \begin{minipage}[t]{4cm}
            \begin{figure}[H]
                \centering                \boxed{\includegraphics[width=\columnwidth]{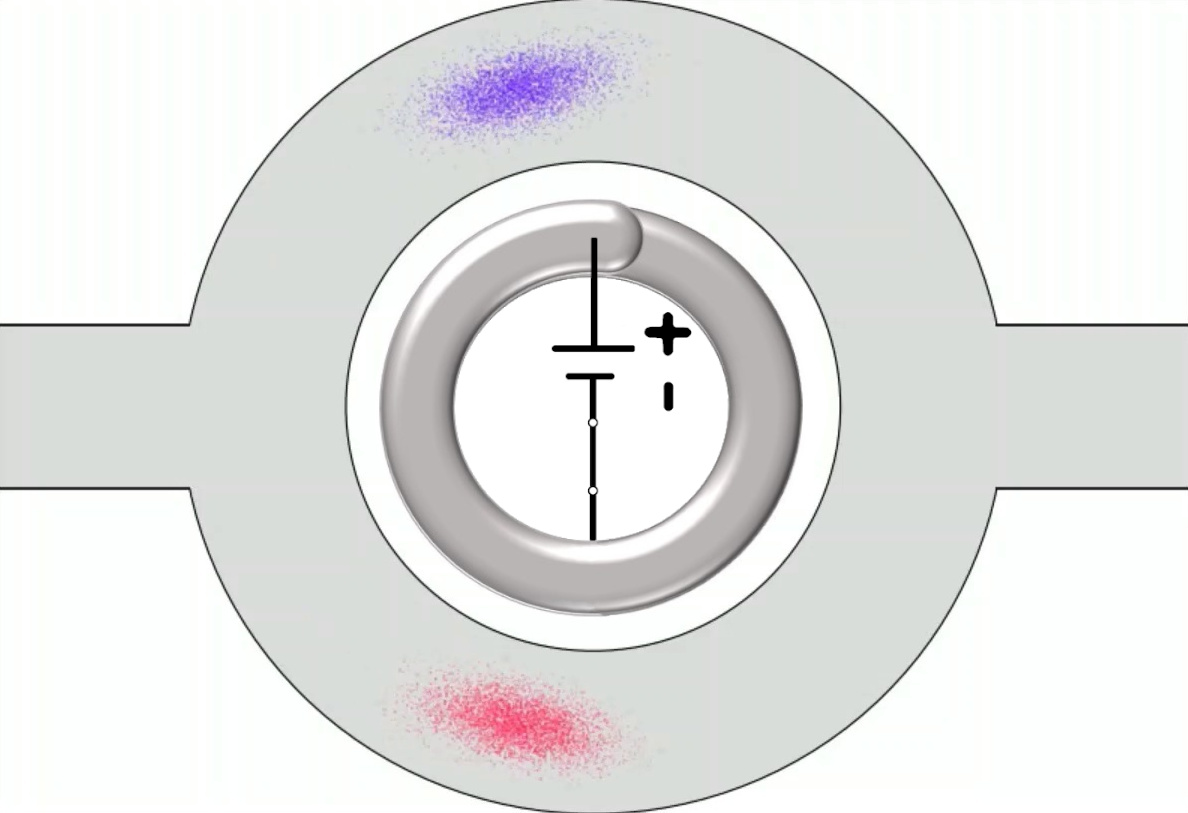}}
                
                \vspace{2mm}(c)
            \end{figure}
        \end{minipage}
        \hfill
        \begin{minipage}[t]{4cm}
            \begin{figure}[H]
                \centering                \boxed{\includegraphics[width=\columnwidth]{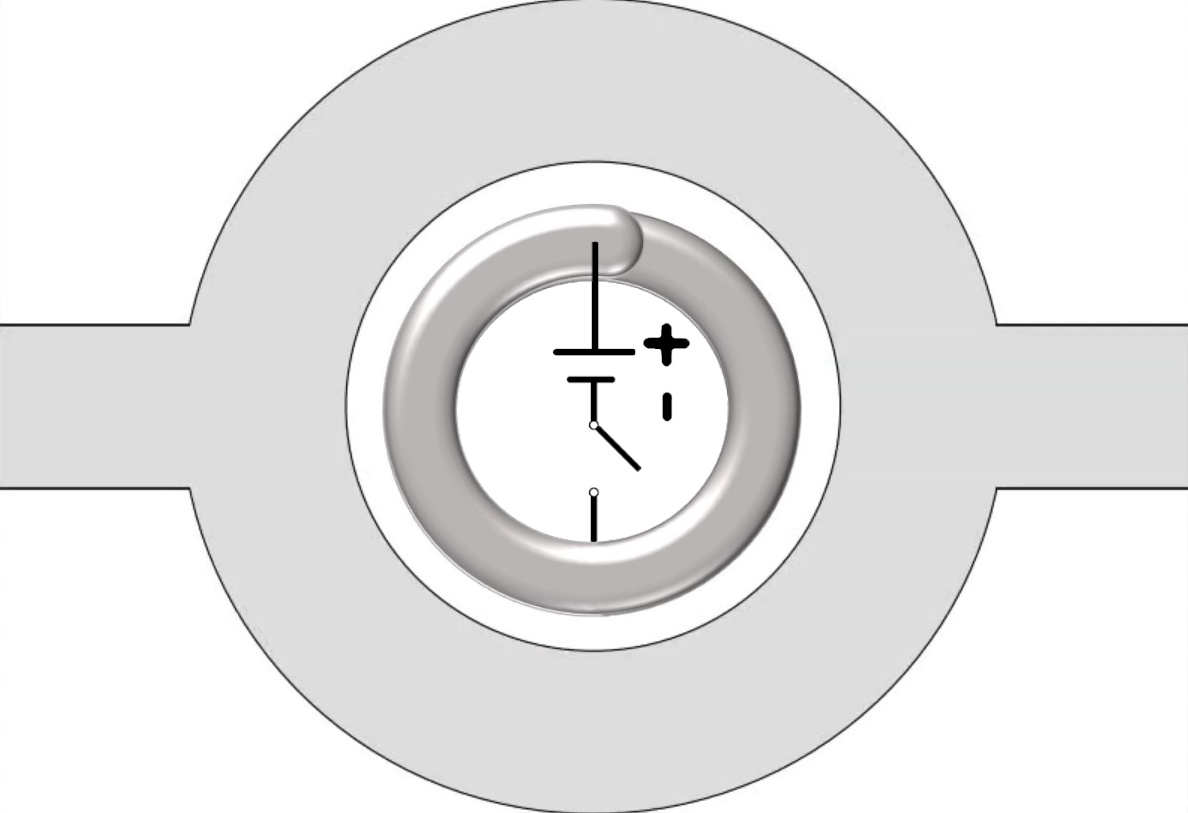}}
                
                \vspace{2mm}(d)
            \end{figure}
        \end{minipage}
    \end{minipage}
    \hfill
    \begin{minipage}[t]{2.24cm}
        \begin{figure}[H]
            \centering
            \includegraphics[width=0.8\columnwidth]{imagenes/color1.png}
        \end{figure}
    \end{minipage}\hspace{2cm}
    \vspace{2mm}
    \captionof{figure}{Graphical representation of a wave packet which evolves over time while travelling through a circular conduit in which centre, a solenoid is placed. Although the phases correspond with a plane wave, in the graphical representation we represent the particle as a localized Gaussian distribution in order to visualize it easier. We see in frame (b) that the probability amplitude split into two (one clockwise and another counterclockwise). Then, in frame (c), the phase of the two components evolve differently, due to the non-zero magnetic vector potential created by the solenoid, until frame (d) when the two components rejoin with a phase difference of $\pi$ radians. Therefore, destructive interference of the two components results and the probability vanishes. The time interval between the frame (a) and the frame (d) is taken as $t=\uptau=4\pi/E_0$.}
    \label{fig:magnetico}
\end{minipage}
\end{figure*}
Once again, it is necessary to solve the eigenvalue equation. In Segal-Bargmann space, it is given by
\begin{eqnarray}
    &T_{SB}[\hat{H}] \phi(z)&=\frac{m}{2}\left(-\frac{1}{2}\left[z^2+\partial^2_z-2z\partial_z-1\right]\right.\nonumber\\
    & &-i\sqrt{2}\frac{q}{m}A(z-\partial_z)\left.+\frac{q^2}{m^2}A^2\right)\phi(z)\nonumber\\
    &&=E\phi(z),
\end{eqnarray}
or
\begin{eqnarray}
    &[&\left.z^2+i2\sqrt{2}\dfrac{qA}{m}z+\partial^2_z-i2\sqrt{2}\dfrac{qA}{m}\partial_z\right.\nonumber\\
    &&-2z\partial_z\left.-\left(1-\dfrac{4E}{m}+2\dfrac{q^2A^2}{m^2}\right)\right]\phi(z)=0.\hspace{8mm}
\end{eqnarray}
The solution of this eigenvalue equation is
\begin{eqnarray}
    &\phi(z)&=C_1\ \text{\large e}^{\frac{z}{2}(i2\sqrt{2}\sqrt{2E/m}+i2\sqrt{2}qA/m+z)}\nonumber\\
    &&+C_2\ \text{\large e}^{\frac{z}{2}(-i2\sqrt{2}\sqrt{2E/m}+i2\sqrt{2}qA/m+z)},\hspace{2mm}
\end{eqnarray}
where $C_1$ and $C_2$ are constants. The Hamiltonian and momentum operator commute ($[\hat{H},\hat{P}]=0$), thus it must exist an eigenfunction basis common to both operators. Applying the momentum operator, we can conclude that
\begin{eqnarray}
\phi_+(z)&=&C_1\ \text{\large e}^{\frac{z}{2}\left(i2\sqrt{2}\left[\sqrt{\frac{2E}{m}}+\frac{qA}{m}\right]+z\right)},\nonumber\\\text{with}&&p=\left[\sqrt{2mE}+qA\right]; \nonumber\\ \phi_-(z)&=&C_2\ \text{\large e}^{\frac{z}{2}\left(i2\sqrt{2}\left[-\sqrt{\frac{2E}{m}}+\frac{qA}{m}\right]+z\right)},\nonumber\\ \text{with}&&p=-\left[\sqrt{2mE}-qA\right].\end{eqnarray}

If the system is initially in an energy eigenstate (with null magnetic vector potential) and in a superposition of momentum states (one clockwise and the other one counterclockwise), then the function that describes the system is

\begin{equation}
    \phi(z,0)=C_1\text{\large e}^{\frac{z}{2}(i2\sqrt{2}p_0/m+z)}+C_2\text{\large e}^{\frac{z}{2}(-i2\sqrt{2}p_0/m+z)},
\end{equation}
where $p_0=\sqrt{2mE_0}$ is the modulus of the momentum vector that the two components have initially.
\newpage
When the solenoid is turned on, both components remain as energy eigenstates, but now each one takes a different energy value. By relating the exponentials arguments, we conclude that 
\begin{eqnarray}
    &p\geq0\Rightarrow E_+=\dfrac{(p_0-qA)^2}{2m},\nonumber\\
    &p\leq0\Rightarrow E_-=\dfrac{(p_0+qA)^2}{2m}.
\end{eqnarray}

Therefore, the system at $t>\uptau$ is described by
\begin{eqnarray}
        &\phi(z,t)&=\text{\large e}^{-iE_+\uptau}\left[C_1\text{\large e}^{\frac{z}{2}(i2\sqrt{2}p_0/m+z)}\nonumber\right.\\
        &&\;\;\;\left.+C_2\text{\large e}^{-i\Delta E\uptau}\text{\large e}^{\frac{z}{2}(-i2\sqrt{2}p_0/m+z)}\right],
\end{eqnarray}
where $\Delta E=2p_0qA/m$. From this expression for $\phi(z,t)$, we can deduce that $\rvert\phi(z,t)\rvert^2\neq \rvert\phi(z,0)\rvert^2$,
if $\Delta E\uptau \neq 2 \pi n$ with $n$ an integer. It means that the probability of measuring the particle at a time $t$, compared to the probability of measuring it initially, is different from $1$. Therefore, the existence of a non-zero magnetic vector potential has a measurable consequence, even though the magnetic field is zero in the region where the particle is propagating.

Moreover, we can compare our result with the result that is obtained in the Schrödinger formalism \cite{Z.3.A.}. If we supposed the same initial conditions in that formalism, then we would obtain that the wave function, at $t>\uptau$, is \footnote{We should remember that the Segal-Bargmann space has been built with dimensionless variables, so the wave function must be expressed in terms of these variables.}
\begin{equation}\label{eq_TFG:resultado_func_onda}
    \psi(s',t)=C\text{\large e}^{-iE_+\uptau}\left[\text{\large e}^{ip_0s'/m}+\text{\large e}^{-i\Delta E\uptau}\text{\large e}^{-ip_0s'/m}\right],
\end{equation}
where $E_+$ and $\Delta E$ are exactly the same as the ones we have obtained in Segal-Bargmann formalism.\\

\vspace{-2mm}By applying the Segal-Bargmann transform to the previous wave function, we obtain
\vspace{-0.3cm}
\begin{eqnarray}
    &\phi_{\psi}(z,t)=&C\sqrt{2}\pi^{1/4}\text{\large e}^{-\frac{{p_0}^2}{2m^2}}\text{\large e}^{-iE_+\uptau}\nonumber\\
    &&\cdot\left[\text{\large e}^{\frac{z}{2}(i2\sqrt{2}p_0/m+z)}+\text{\large e}^{-i\Delta E\uptau}\right.\nonumber\\
    &&\cdot\left.\text{\large e}^{\frac{z}{2}(-i2\sqrt{2}p_0/m+z)}\right].
\end{eqnarray}

It should be noted that taking $C_1=C_2=C\sqrt{2}\pi^{1/4}\text{exp}(-p_0^2/2m^2)$, the results obtained computing all the solutions using only the Segal-Bargmann space formalism are identical to the ones obtained through the well-known solutions in position representation.

On the other hand, in phase space, the Hamiltonian operator described by \eqref{eq_TFG:ham_A} is expressed as follows
\newpage
\begin{widetext}
\begin{eqnarray}
    &T_W^{-1}[\hat{H}]\star&=\frac{1}{2m}\left[(p-\frac{i}{2}\partial_s)^2-2qA(p-\frac{i}{2}\partial_s)+q^2A^2\right]\nonumber\\
    &&=\frac{1}{2m}\left[p^2-\frac{1}{4}\partial^2_s-ip\partial_s-2qAp+iqA\partial_s+q^2A^2\right].
\end{eqnarray}
\end{widetext}
The eigenvalue equation is 
\begin{eqnarray}
    &\left[(p-qA)^2-2mE-\frac{1}{4}\partial^2_s\right.\nonumber\\
    &\hspace{1.1cm}\left.-i(p-qA)\partial_s\right]W(s,p)=0.
\end{eqnarray}
By using the same procedure that we used for the free-particle case, we reach the following solutions, which are eigenstates with energy $E$ and momentum $p$.
\vspace{-0.4cm}
\begin{eqnarray}
W_+(s,p)&=&C_1\ \delta(p-[\sqrt{2mE}+qA]),\nonumber\\
\text{with}&&p=\left[\sqrt{2mE}+qA\right]; \nonumber\\ W_-(s,p)&=&C_2\ \delta(p+[\sqrt{2mE}-qA]),\nonumber\\ \text{with}&&p=-\left[\sqrt{2mE}-qA\right].\end{eqnarray}

If the system is initially in an energy eigenstate (with null magnetic vector potential) and in a superposition of momentum states (one clockwise and the other one counterclockwise), then the Wigner function that describes the system will be the sum of the terms associated with the free-particle solutions clockwise and counterclockwise, $W_{1,1}$ and $W_{2,2}$, together with the sum of the cross-terms $W_{1,2}$ y $W_{2,1}$.

It is easy to verify that $W_{1,2}=W_{2,1}^*=\text{exp}(i2p_0s)\delta(p)$ fulfil the eigenvalue equation \eqref{eq_TFG:no_diag_wigner}, where $p_0=\sqrt{2mE_0}$ is again the modulus of the initial momentum vector of the particle.

Hence, the initial Wigner function is
\begin{eqnarray}
    &&W(s,p,0)=  C_{1,1}\delta(p-p_0)+C_{2,2}\delta(p+p_0)\nonumber\\
    &&\;\;\;\;\;\;\;\;\;\;\;\,
    +C_{1,2}\text{\large e}^{i2p_0s}\delta(p)+C_{2,1}\text{\large e}^{-i2p_0s}\delta(p).\hspace{8mm}
\end{eqnarray}

When the solenoid is turned on, the above eigenfunctions remain as energy eigenstates, but now each one of them takes a different energy value. By relating the delta arguments of $W_{1,1}$ and $W_{2,2}$ with the energy and momentum eigenstates, we conclude that
\begin{eqnarray}
    &p\geq0\Rightarrow E_+=\dfrac{(p_0-qA)^2}{2m},\nonumber\\
    &p\leq0\Rightarrow E_-=\dfrac{(p_0+qA)^2}{2m}.
\end{eqnarray}
Therefore, by using Equations \eqref{eq_TFG:ev_wigner} and \eqref{eq_TFG:no_diag_wigner}, we deduce that the system at $t>\uptau$, is described by
\begin{eqnarray}\label{eq_TFG:Wigner_fases}
    &W(s,p,t)&=C_{1,1}\delta(p-p_0)+C_{2,2}\delta(p+p_0)\nonumber\\
    &&\;\;\;+C_{1,2}\text{\large e}^{-i(\Delta E\uptau-2p_0s)}\delta(p)\nonumber\\
    &&\;\;\;+C_{2,1}\text{\large e}^{i(\Delta E\uptau-2p_0s)}\delta(p),
\end{eqnarray}
\noindent where $\Delta E=2p_0qA/m$. It should be noted that this phase difference is the same as the phase difference obtained in Segal-Bargmann space, which was the same as the one predicted by the Schrödinger formalism.

Also in this case, it is easy to deduce from the expression of $W(s,p,t)$, that $W(s,p,t)\neq W(s,p,0)$, when $\Delta E\uptau \neq 2 \pi n$ with $n$ an integer. Therefore, the probability of measuring the particle at a time $t$, compared to the initial probability of measuring it, has been modified by the existence of a non-zero magnetic vector potential even the magnetic field is null over the area where the particle is propagating.

Moreover, we can compare our result with the result obtained in the Schrödinger formalism \cite{Z.3.A.}. As we had seen in the previous section, the wave function that is obtained in Schrödinger formalism is given by Equation \eqref{eq_TFG:resultado_func_onda}. By using Equation \eqref{eq_TFG:Wigner_pos}, the Wigner function reads
\begin{eqnarray}\label{eq_TFG:Wigner_MC}
    &&W_{\psi}(s,p,t)=C^2\left(\delta(p-p_0)+\delta(p+p_0)\right.
    \nonumber\\
    &&+\left.\left[\text{\large e}^{-i(\Delta E\uptau-2p_0s)}\delta(p)+\text{\large e}^{(i\Delta E\uptau-2p_0s)}\delta(p)\right]\right).\hspace{8mm}
\end{eqnarray}
We could note that taking $C_{i,j}=C^2$, the solution obtained by using exclusively the phase-space formalism of quantum mechanics is identical to the one obtained by transforming the well-known solution in position representation into a Wigner function in phase-space formalism.

With the aim to visualize the Aharonov-Bohm effect with non-zero magnetic vector potential more easily; we build a graphical representation making use of the results that we have obtained within the phase-space and the Segal-Bargmann space formalisms. Once again, in that graphical representation, both the spatial probability of finding the particle and the phase of the particle are represented through colours. Although the results correspond with a plane wave, in the graphical representation we represent the particle as a localized Gaussian distribution in order to visualize it easier.

As it can be seen in Fig. \ref{fig:magnetico}a, we begin with a circular conduit in which centre, a solenoid is placed. When we connect that solenoid (with $\rvert\Vec{A}\rvert=p_0/16q$ during a time interval of $t=\uptau=4\pi/E_0$), the components that travel clockwise and counterclockwise acquire different phases; thereby, when we disconnect the solenoid, the phase difference between those components is $\pi$ radians (Fig. \ref{fig:magnetico}b and Fig. \ref{fig:magnetico}c). It means that the probability of finding the particle, for $t\geq\uptau$, vanishes (Fig. \ref{fig:magnetico}d).

\subsection{General Aharonov-Bohm effect}\label{subsec4.4}

Finally, we will find an expression for general systems affected by the Aharonov-Bohm effect, without supposing a specific case of study \cite{Z.3.A.}.

To that end, we make use of a one-dimensional system, where electric and magnetic fields are null over the accessible region, and the magnetic vector potential does not change over time. The latter condition implies that the magnetic vector potential depends only on position and the electric potential depends only on time ($E=-\partial \varphi /\partial x-\partial A/\partial t$). This case is described by the hamiltonian \cite{I.J-S}:
\begin{equation}
    \hat{H}=\frac{1}{2m}(\hat{P}-qA(\hat{x}))^2+q\varphi(t).
\end{equation}

We obtain the global phase in each of the spaces that we have been considering in this work by proposing an expression that is solution of the corresponding evolution equations. In the Schrödinger formalism, the proposed solution for a particle that begins in $(x_0,t_0)$, is the operator $\hat{M}(\hat{x})=\text{exp}(iq[\int_{x_0}^{\hat{x}} dr\ A-\int_{t_0}^tdt'\ \varphi])$ acting on the state of a free particle $\ket{\psi_0}$.

In Segal-Bargmann space, the operator used to build the solution of the evolution equation is the Segal-Bargmann transform of the previous operator
\begin{equation}
    T_{SB}[\hat{M}]=\text{\large e}^{i[\frac{q}{m\sqrt{2}}\int_{\sqrt{2}mx_0}^{z+\partial_z}dr\ A-q\int_{t_0}^tdt'\ \varphi]}.
\end{equation}
It can be seen that this operator fulfils one useful property \footnote{This property may actually be posed as the commutation relation $[\hat{P}-qA,\hat{M}]$, which implies it is fulfilled in all formalisms.}
\begin{widetext2}
\begin{equation}
    T_{SB}[\hat{P}-qA]\cdot T_{SB}[\hat{M}]=\text{\large e}^{i[\frac{q}{m\sqrt{2}}\int_{\sqrt{2}mx_0}^{z+\partial_z}dr\ A-q\int_{t_0}^tdt' \varphi]}\cdot\left(\frac{im}{\sqrt{2}}(z-\partial_z)-\frac{qA}{2}\right)=T_{SB}[\hat{M}]\cdot T_{SB}[\hat{P}].
\end{equation}
\end{widetext2}
Taking this result into account, we can analyse the evolution Equation \eqref{eq_TFG:ev_temp_segal} with the solution $T_{SB}[\hat{M}]\phi_0(z)$ (Where $\phi_0(z)$ is the solution of a free particle and $H_0$ is its Hamiltonian):

\begin{eqnarray}
    &&
    T_{SB}[\hat{M}]\left(i\frac{\partial \phi_0}{\partial t}+q\varphi \phi_0\right)\nonumber\\
  &&\;\;\;\;\;\;
  =\left(\frac{1}{2m}T_{SB}[\hat{P}-qA]^2+q\varphi\right)\cdot T_{SB}[\hat{M}]\phi_0\nonumber\\
    &&\;\;\;\;\;\;
    =T_{SB}[\hat{M}]\cdot \left(\frac{1}{2m}T_{SB}[\hat{P}]^2+q\varphi\right)\phi_0\nonumber\\
    &&\;\;\;\;\;\;
    =T_{SB}[\hat{M}]\cdot \left(T_{SB}[\hat{H_0}]\phi_0+q\varphi\phi_0\right).
\end{eqnarray}

Therefore, it is proved that the proposed solution fulfils the evolution equation.

In phase space, the operator posed to generate the solution of the evolution equation is the inverse Weyl transform of $\hat{M}$
\begin{equation}
    T_W^{-1}[\hat{M}]=\text{\large e}^{iq[\int_{x_0}^{x}dr\ A-\int_{t_0}^tdt'\ \varphi]}.
\end{equation}

It can be seen that this operator fulfils the property

\begin{widetext}
\begin{equation}
    (p-qA)\star T_W^{-1}[\hat{M}]=\text{\large e}^{iq[\int_{x_0}^{x}dr\ A-\int_{t_0}^tdt'\ \varphi]}\left(p-\frac{qA}{2}\right)=T_W^{-1}[M]\star p.
\end{equation}
\end{widetext}
As the Wigner function do not show global phases,  it is not possible to obtain the form of the global phase directly from the evolution equation within such a formalism. However, we can pose a system in which the wave function of a particle is split into two, one part passes through a zone with non-zero electromagnetic potentials and later both parts recombine. As a consequence of that, the phase difference between both parts would be the global phase acquired by the part of the wave function that is affected by the Aharonov-Bohm effect. In that case, the only non-trivial equation of evolution given by Equation \eqref{eq_TFG:ev_wigner} is that corresponding to $W_{1,2}$; and the solution we propose is $T_W^{-1}[\hat{M}]\star W_0(x,p)$ (with the subscript $1$ referring to the wave function that is affected by the Aharonov-Bohm effect and the subscripts $2$ and $0$ referring to the free particle).

\begin{eqnarray}
        &&T_W^{-1}[\hat{M}]\star\left(i\frac{\partial W_0}{\partial t}+q\varphi W_0\right)=\nonumber\\
        &&\;\;\;\;\;\;
        =H\star T_W^{-1}[\hat{M}]\star W_0-T_W^{-1}[\hat{M}]\star W_0\star H
        \nonumber\\
        & &\;\;\;\;\;\;
        =\left(\frac{(p-qA)^2}{2m}+q\varphi\right)\star T_W^{-1}[\hat{M}]\star W_0\nonumber\\
        &&\;\;\;\;\;\;\;\;\;\;
        -T_W^{-1}[\hat{M}]\star W_0\star H_0
        \nonumber\\
        & &\;\;\;\;\;\;
        =T_W^{-1}[\hat{M}]\star \left(\frac{p^2}{2m}+q\varphi\right)\star W_0\nonumber\\
        &&\;\;\;\;\;\;\;\;\;\;
        -T_W^{-1}[\hat{M}]\star W_0\star H_0
        \nonumber\\
        & &\;\;\;\;\;\;
        =T_W^{-1}[\hat{M}]\star \left([H_0,W_0]_M+q\varphi W_0\right)\,.
\end{eqnarray}

Therefore, it is proved that the proposed solution fulfils the evolution equation.

In each formalism, the proposed solution has been an operator acting on the solution of the free particle. Repeating the procedure, we then obtain the same solution together with a phase. That is the global phase acquired by the particle due to Aharonov-Bohm effect
\begin{equation}
    \theta=q\left[\int_{x_0}^{x}dr\ A-\int_{t_0}^tdt'\ \varphi\right].
\end{equation}

By extending the previous results to three dimensions and rewriting it in covariant notation, the global phase acquired by the particle becomes
\begin{equation}
    \label{eq_TFG:notacion_covar}
    \theta=q\int_{C}dx_\mu\ A^\mu,
\end{equation}

\noindent where $x^\mu$ are the space-time coordinates, $A^\mu$ is the electromagnetic four-potential and $C$ is the curve that the charged particle describes. In addition, from Equation \eqref{eq_TFG:notacion_covar}, it can be seen that a gauge transformation adds a global phase to the wave function; so observables are invariant under gauge transformation.

Although we have rewritten the expression of the global phase in a more compact way by using covariant notation, it is important to note that we have demonstrated its validity only for non-relativistic systems.

\subsection{Aharonov-Bohm effect with a density operator}

Suppose a superposition of two states, one with well-defined position, $q_0$, and another one with well-defined momentum, $p_0$; but with undetermined coherence. The density operator of such state is thus
\begin{equation}
    \hat{\rho}=\alpha\ket{x_0}\bra{x_0}+\gamma\ket{x_0}\bra{p_0}+\gamma^*\ket{p_0}\bra{x_0}+\beta\ket{p_0}\bra{p_0},
\end{equation}
with $\alpha,\beta\in \mathbb{R}$ and $\gamma\in \mathbb{C}$. If $\hat{\rho}$ is a density operator, it must be self-adjoint and positive semi-definite\footnote{We do not include the condition $\text{Tr}[\hat{\rho}]=1$ because it is related to the norm of the state. Once again, we are taking a non-normalizable state as a simple limit of a normalizable one}. For the last condition to be fulfilled, it is necessary that $\alpha\beta-\rvert\gamma\rvert^2\geq0$ with $\alpha,\beta>0$. We can see that if $\gamma=0$, there is a complete incoherent superposition, but if $\rvert \gamma\rvert^2=\alpha\beta$, there is a complete coherent superposition. With the aim of simplify the notation, that density operator will be expressed as $\hat{\rho}=\alpha \hat{\rho}_{11}+\gamma \hat{\rho}_{12}+\gamma^* \hat{\rho}_{21}+\beta \hat{\rho}_{22}$.

By using the inverse Weyl transform \eqref{eq_TFG:TW_pos} we obtain the following Wigner functions\footnote{Note that we have removed a global factor $1/2\pi$ since it is irrelevant due to the non-normalizability of the Wigner function.}
\begin{eqnarray}\label{eq:in_cond_mixto}
    W(x,p)&=& \alpha W_{11}+\gamma W_{12}+\gamma^* W_{21}+\beta W_{22}.\nonumber\\
    W_{11}(x,p)&=&\delta(x-x_0).\nonumber\\
    W_{22}(x,p)&=&\delta(p-p_0).\nonumber\\
    W_{12}(x,p)&=&\frac{2\gamma}{\sqrt{2\pi}}\text{\large e}^{i\left[2p(x-x_0)-p_0(2x-x_0)\right]}.\nonumber \\
    W_{21}(x,p)&=&W_{12}(x,p)^*.
\end{eqnarray}

Let us now assume a system similar to that of section \ref{subsec4.2}. At time $t=0$, the density operator $\hat{\rho}$ describes the quantum state with $\ket{x_0}$ in one conduit and $\ket{p_0}$ in the other one. At the same time, the electric device is turned on, and it remains on until $t=\uptau$ when it is turned off.

We can calculate the time evolution of each Wigner function $W_{ij}$, $i,j=1,2$, by using \eqref{eq_TFG:ev_wigner}
\begin{equation}\label{eq:ev_wigner_mixto}
    \frac{\partial W_{ij}}{\partial t}=-i\left[H_i\star W_{ij}-W_{ij}\star H_j\right],
\end{equation}
where $H_1$ is the hamiltonian associated with the conduit where is $\ket{x_0}$ and $H_2$ is the hamiltonian associated with the conduit where is $\ket{p_0}$.

When $i=j$, \eqref{eq:ev_wigner_mixto} becomes
\begin{equation}
    \frac{\partial W_{ii}}{\partial t}+ \frac{p}{m}\frac{\partial W_{ii}}{\partial x}=0.
\end{equation}

The solution of this first-order partial differential equation (PDE) is
\begin{equation}
    W_{ii}(x,p,t)=W(\xi,p),
\end{equation}
where $\xi=t-mx/p$. By using the initial conditions $W_{11}(x,p,0)$ and $W_{22}(x,p,0)$ of \eqref{eq:in_cond_mixto}, we obtain
\begin{eqnarray}
    W_{11}(x,p,t)&=&\delta\left(x-x_0-pt/m 
 \right).\nonumber\\
    W_{22}(x,p,t)&=&\delta\left(p-p_0\right).
\end{eqnarray}

When $i\neq j$ and $0\leq t<\uptau$, \eqref{eq:ev_wigner_mixto} becomes
\begin{equation}
    \frac{\partial W_{12}}{\partial t}+ \frac{p}{m}\frac{\partial W_{12}}{\partial x}=iq\Delta \varphi W_{12}.
\end{equation}

The equation of $W_{21}$ could be derived from the later equation by taking the complex conjugate. The solution of this first-order PDE is
\begin{equation}\label{eq:w12_mixto}
    W_{12}(x,p,t)=W(\xi,p)\text{\large e}^{iq\Delta \varphi \cdot t}.
\end{equation}
By using the initial conditions $W_{12}(x,p,0)$ and $W_{21}(x,p,0)$ of \eqref{eq:in_cond_mixto}, we obtain
\begin{eqnarray}
    W_{12}(x,p,t)&=&\frac{2\rvert\gamma\rvert}{\sqrt{2\pi}}\text{\large e}^{i\theta(x,p,t)}.\nonumber\\
    W_{21}(x,p,t)&=&\frac{2\rvert\gamma\rvert}{\sqrt{2\pi}}\text{\large e}^{-i\theta(x,p,t)}.\nonumber\\
    \theta(x,p,t)&\equiv&q\Delta \varphi \cdot t+\text{Arg}(\gamma)+2p\left(x-x_0-\frac{p}{m}t\right)\nonumber\\
    &&-p_0\left(2x-x_0-\frac{2p}{m}t\right).
\end{eqnarray}

When $t\geq \uptau$, $H_1$ becomes equal to $H_2$, so the exponential of \eqref{eq:w12_mixto} vanished. The Wigner function that describes the system at $t\geq\uptau$ is thus
\begin{eqnarray}
    W(x,p,t)&=&\alpha\delta\left(x-x_0-pt/m\right)+\beta\delta\left(p-p_0\right)\nonumber\\
    &&+\frac{4\rvert\gamma\rvert}{\sqrt{2\pi}}\cos(\theta(x,p,t)).\nonumber\\
    \theta(x,p,t)&=& q\Delta \varphi \cdot\uptau+\text{Arg}(\gamma)+2p\left(x-x_0-\frac{p}{m}t\right)\nonumber\\
    &&-p_0\left(2x-x_0-\frac{2p}{m}t\right).
\end{eqnarray}

\begin{figure*}[ht]
\begin{minipage}[t]{\textwidth}
    \begin{minipage}[t]{\textwidth}
    \centering
        \hspace{1cm}\begin{minipage}[t]{0.4\columnwidth}
          \begin{figure}[H]
            \centering
            \includegraphics[width=\textwidth]{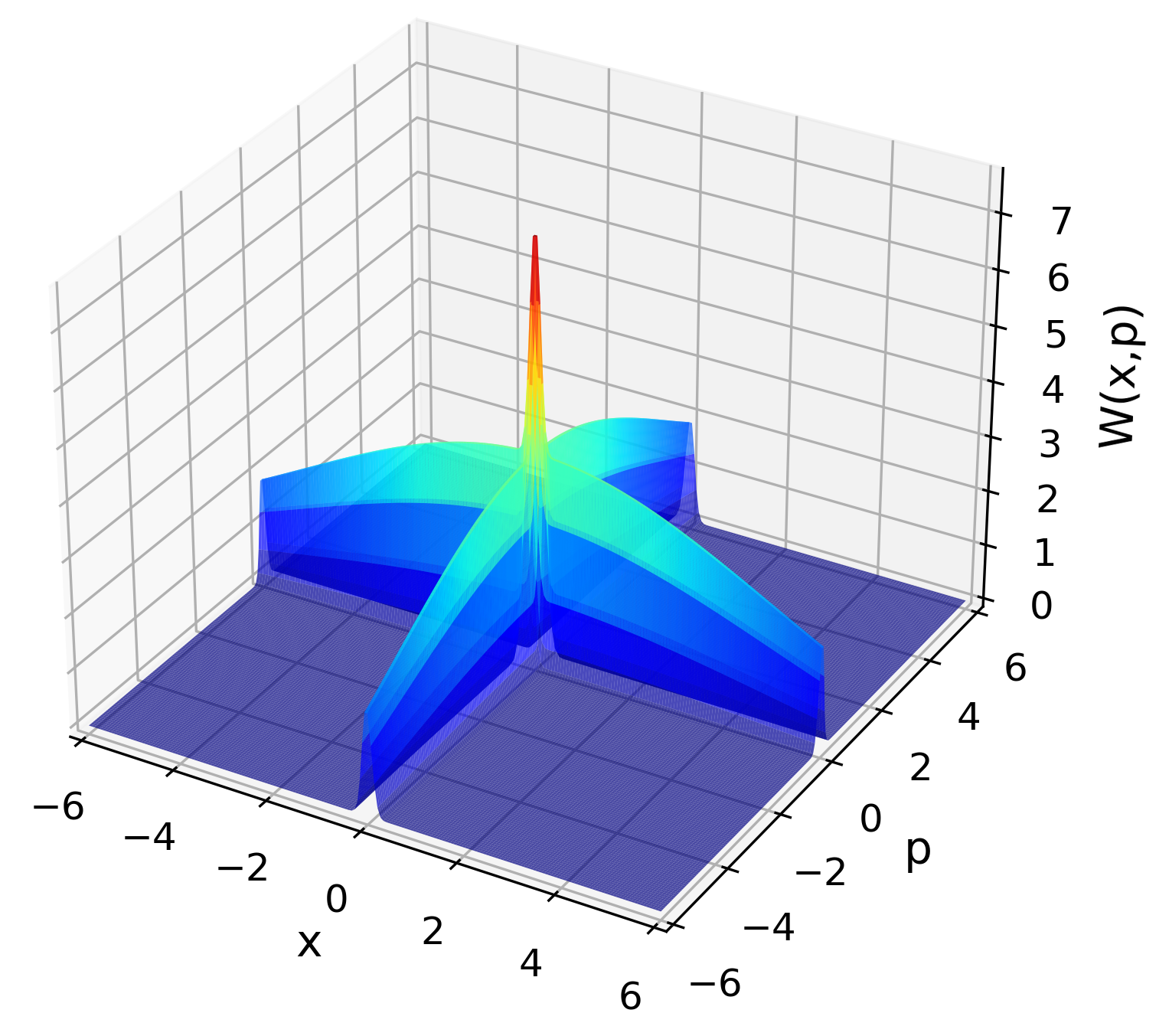}
            \end{figure}
            \vspace{-5mm}\centering (a1)
        \end{minipage}
        \hfill
        \begin{minipage}[t]{0.4\columnwidth}
          \begin{figure}[H]
            \centering
            \includegraphics[width=\columnwidth]{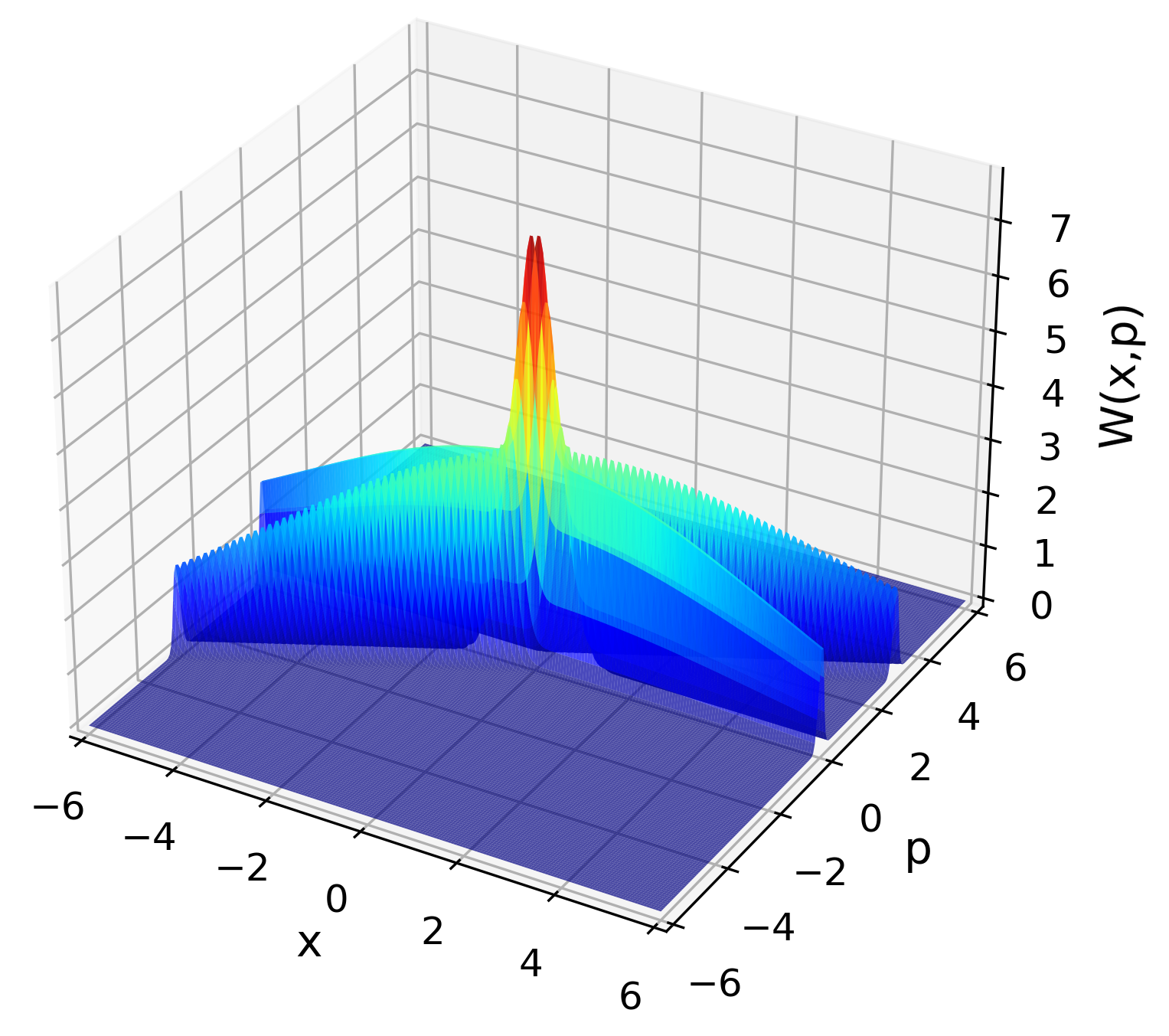}
            \end{figure}
            \vspace{-5mm}\centering (b1)
        \end{minipage}\hspace{1cm}
    \end{minipage}
    \begin{minipage}[t]{\textwidth}
    \centering
        \hspace{1cm}\begin{minipage}[t]{0.4\columnwidth}
            \begin{figure}[H]
            \centering
            \includegraphics[width=\columnwidth]{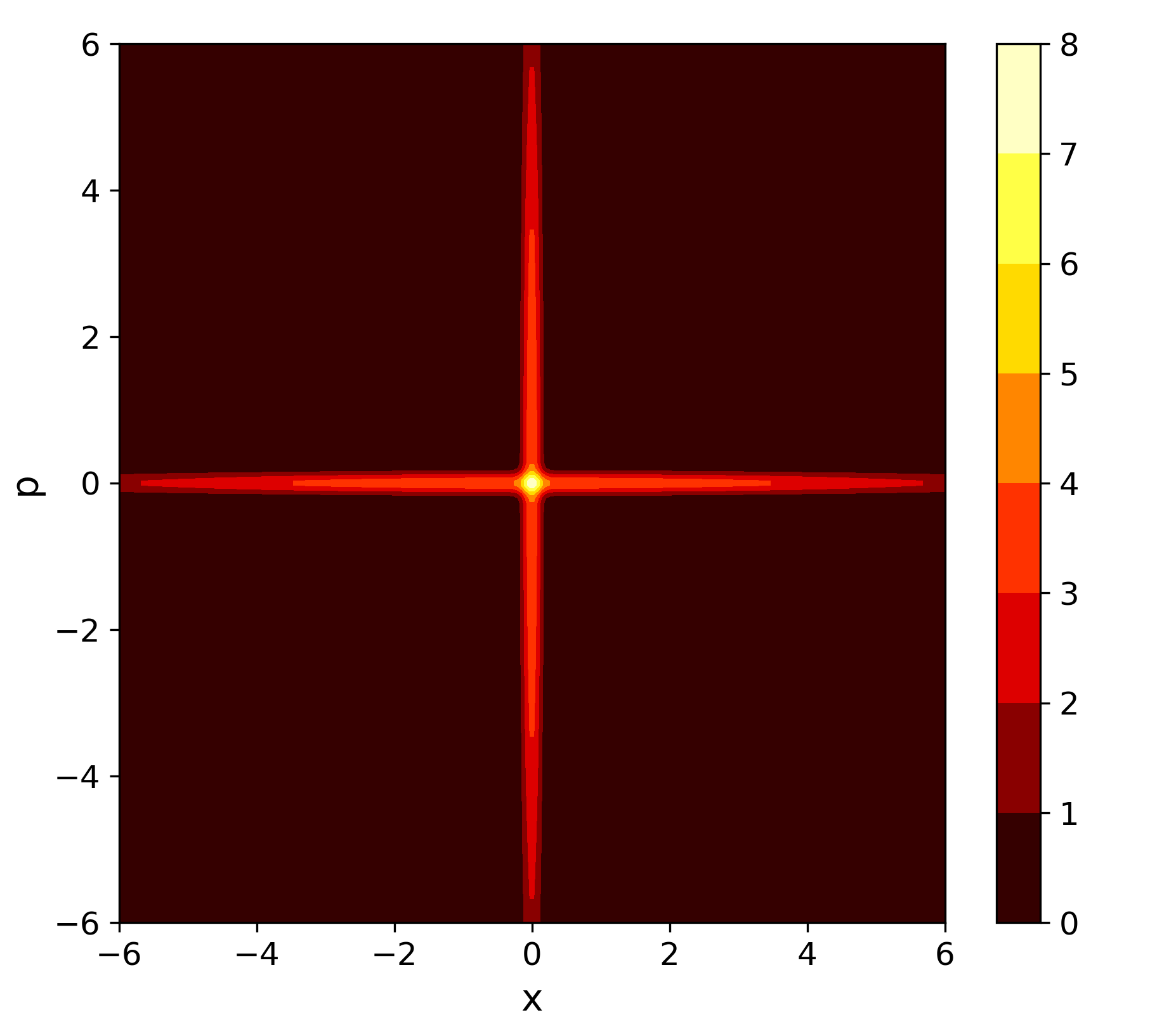}
            \end{figure}
            \vspace{-5mm}\centering (a2)
        \end{minipage}
        \hfill
        \begin{minipage}[t]{0.4\columnwidth}
            \begin{figure}[H]
            \centering
            \includegraphics[width=\columnwidth]{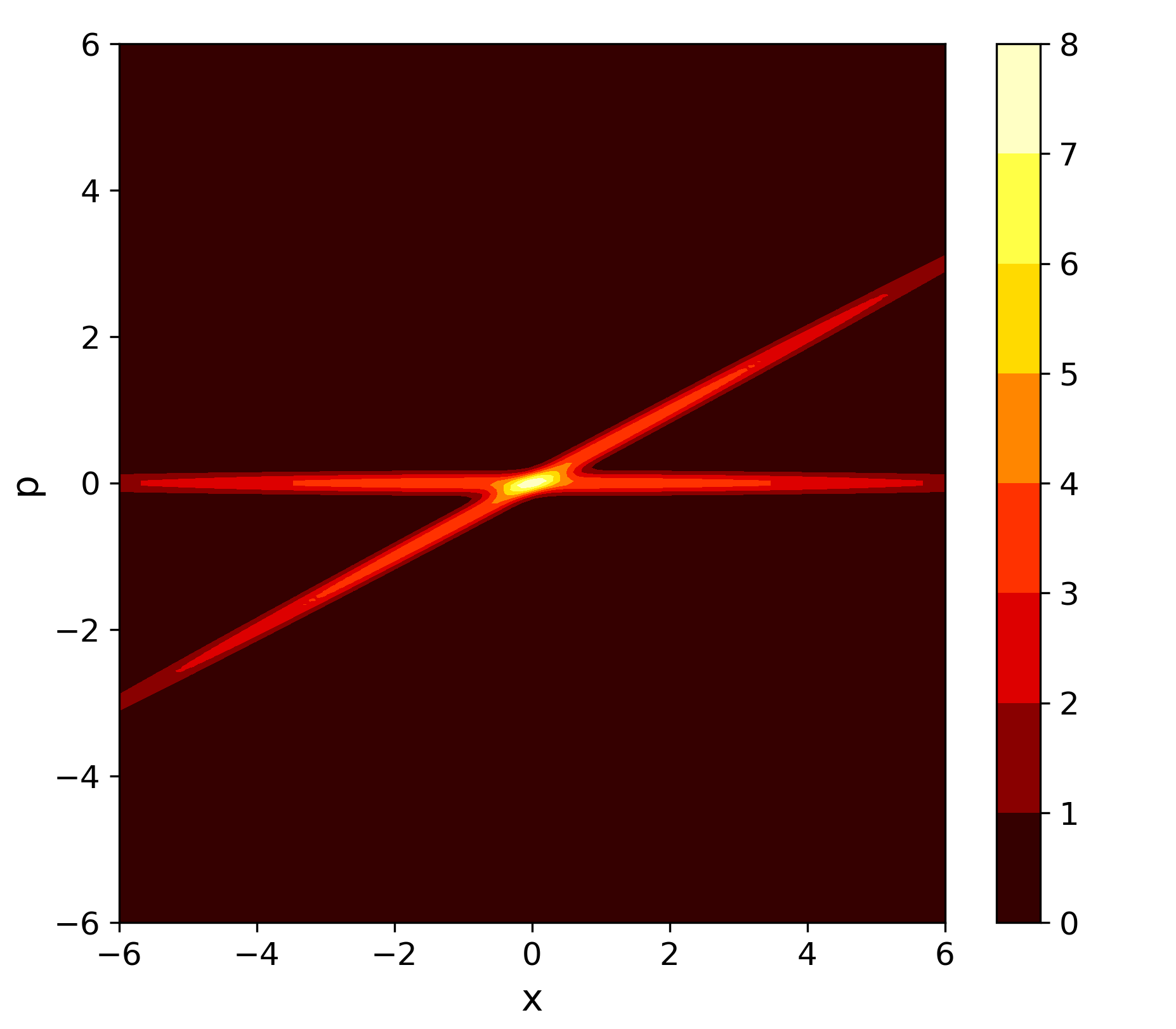}
            \end{figure}
            \vspace{-5mm}\centering (b2)
        \end{minipage}\hspace{1cm}
    \end{minipage}
\vspace{2mm}
\captionof{figure}{Non-normalized Wigner function of the density operator with $\gamma=0$ and $\alpha=\beta=1$. For simplicity, we also set $x_0=p_0=0$, $\Delta\varphi=0$ and $m=1$ (so the phase-space variables and time become dimensionless). For the representation, we do not use delta functions but Gaussian distributions: $\text{exp}(-2\sigma^2p^2-x^2/2\sigma^2)/\sqrt{2\pi\sigma^2}$ with $\sigma=0.1$, thus it can be seen in the figure. On the first row we represent the Wigner function in a 3-D plot, while on the second row it is represented in a 2-D plot whose colour denotes the Wigner function value on that phase-space point. Furthermore, the first column corresponds to the initial time $t=0$, and the second column corresponds to a large time $t=2$.}
\label{fig:incoh}
\end{minipage}
\end{figure*}

\begin{figure*}[ht]
\begin{minipage}[t]{\textwidth}
    \begin{minipage}[t]{\textwidth}
    \centering
        \hspace{1cm}\begin{minipage}[t]{0.4\columnwidth}
          \begin{figure}[H]
            \centering
            \includegraphics[width=\textwidth]{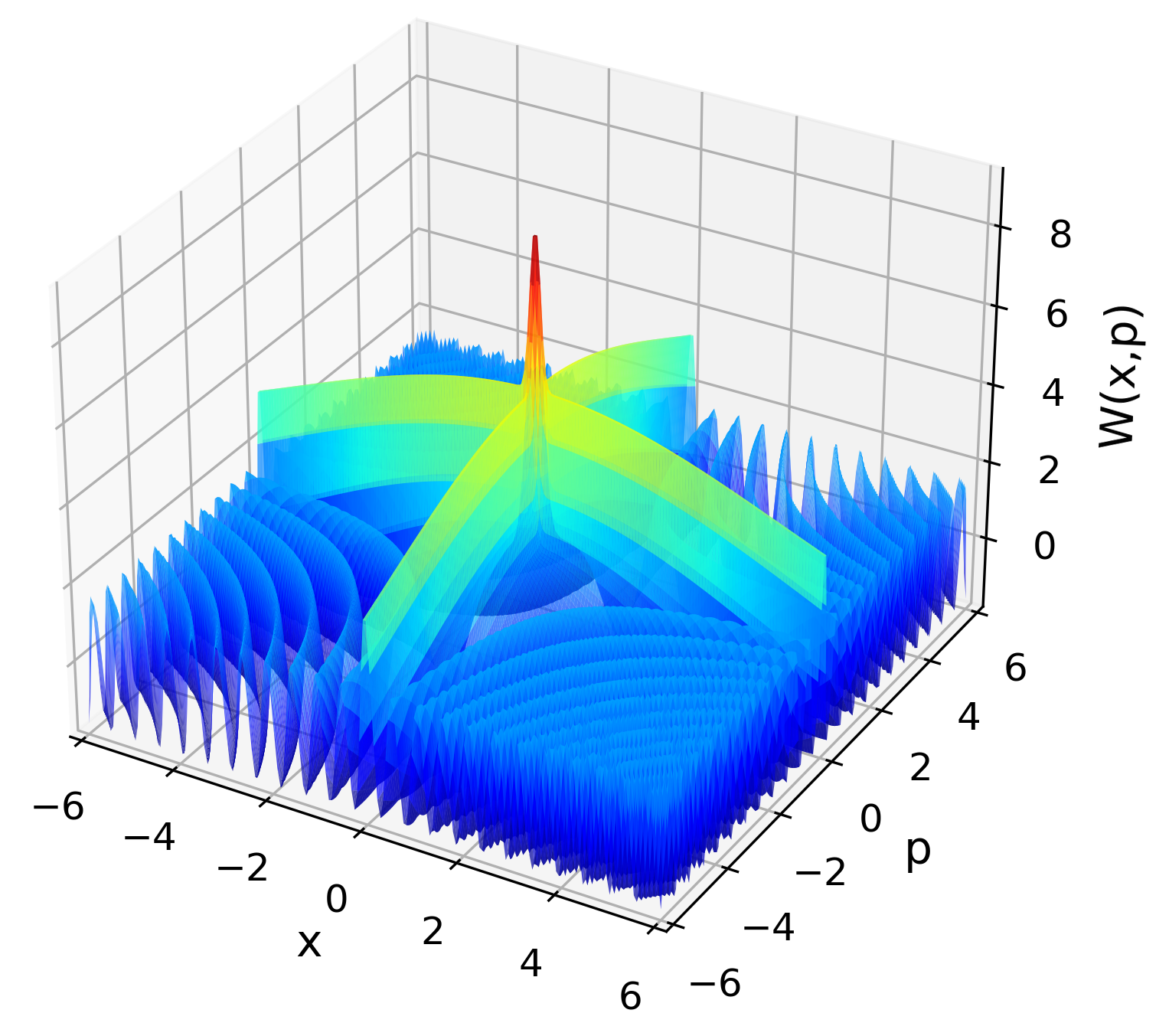}
            \end{figure}
            \vspace{-5mm}\centering (a1)
        \end{minipage}
        \hfill
        \begin{minipage}[t]{0.4\columnwidth}
          \begin{figure}[H]
            \centering
            \includegraphics[width=\columnwidth]{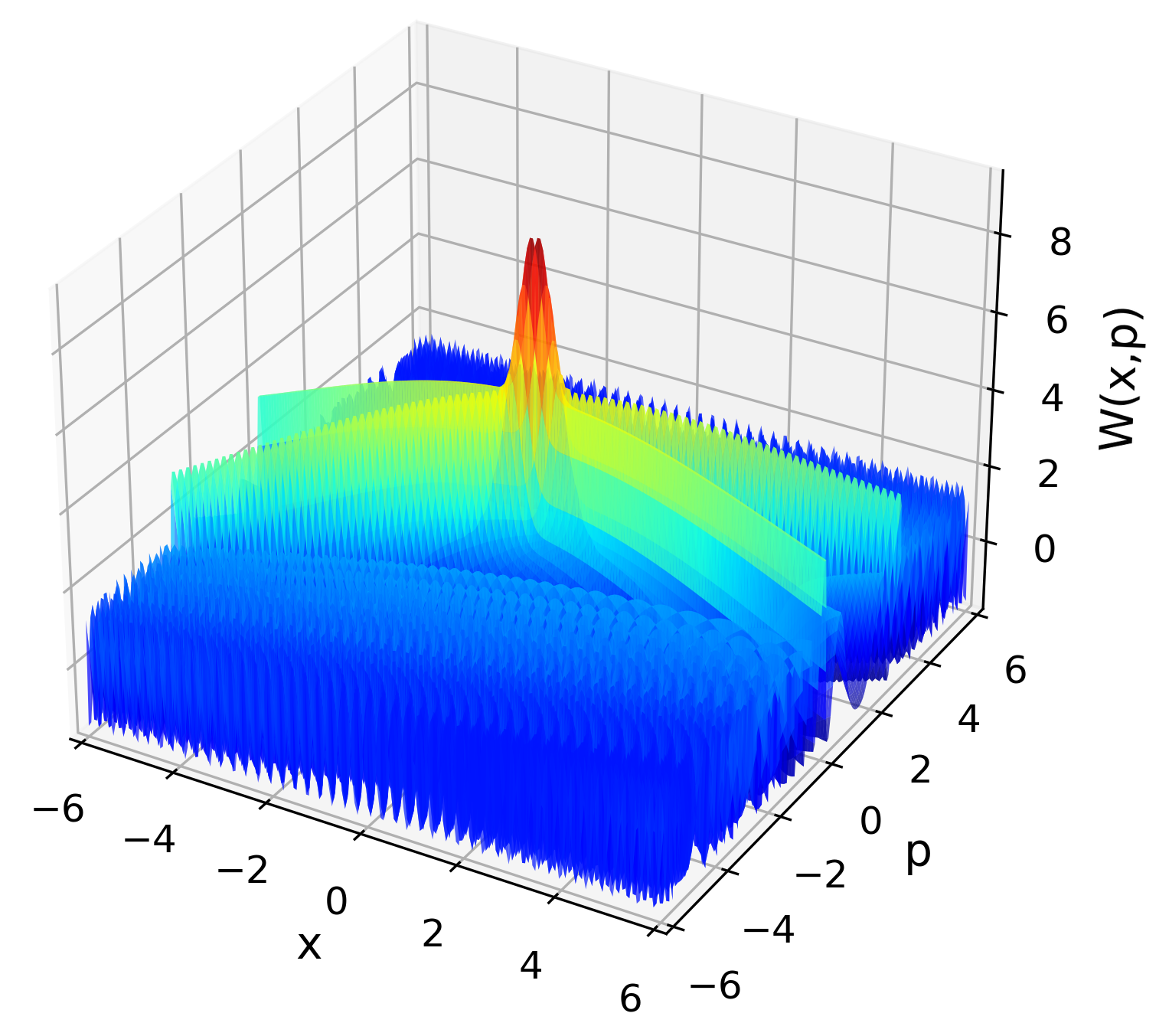}
            \end{figure}
            \vspace{-5mm}\centering (b1)
        \end{minipage}\hspace{1cm}
    \end{minipage}
    \begin{minipage}[t]{\textwidth}
    \centering
        \hspace{1cm}\begin{minipage}[t]{0.4\columnwidth}
            \begin{figure}[H]
            \centering
            \includegraphics[width=\columnwidth]{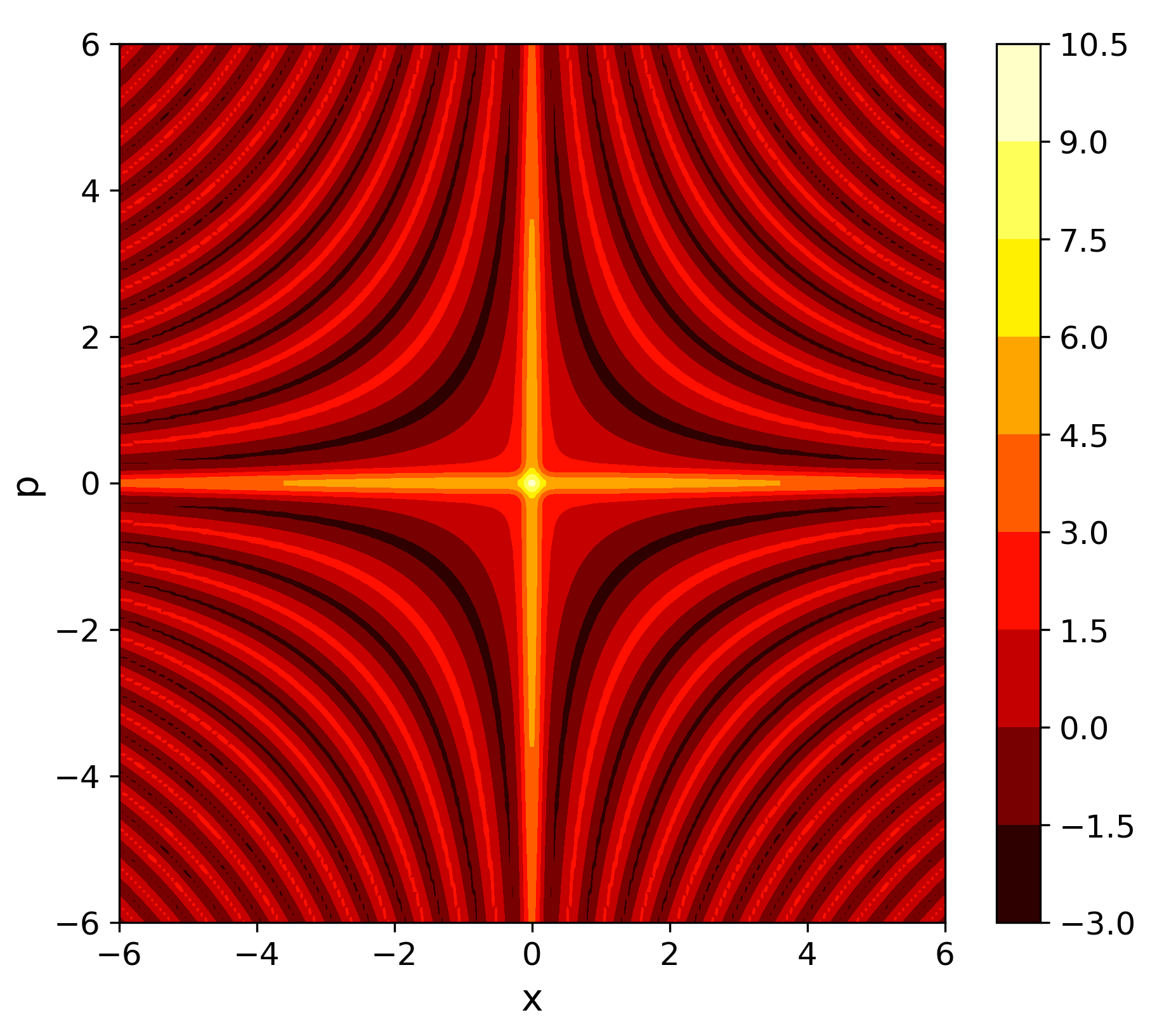}
            \end{figure}
            \vspace{-5mm}\centering (a2)
        \end{minipage}
        \hfill
        \begin{minipage}[t]{0.4\columnwidth}
            \begin{figure}[H]
            \centering
            \includegraphics[width=\columnwidth]{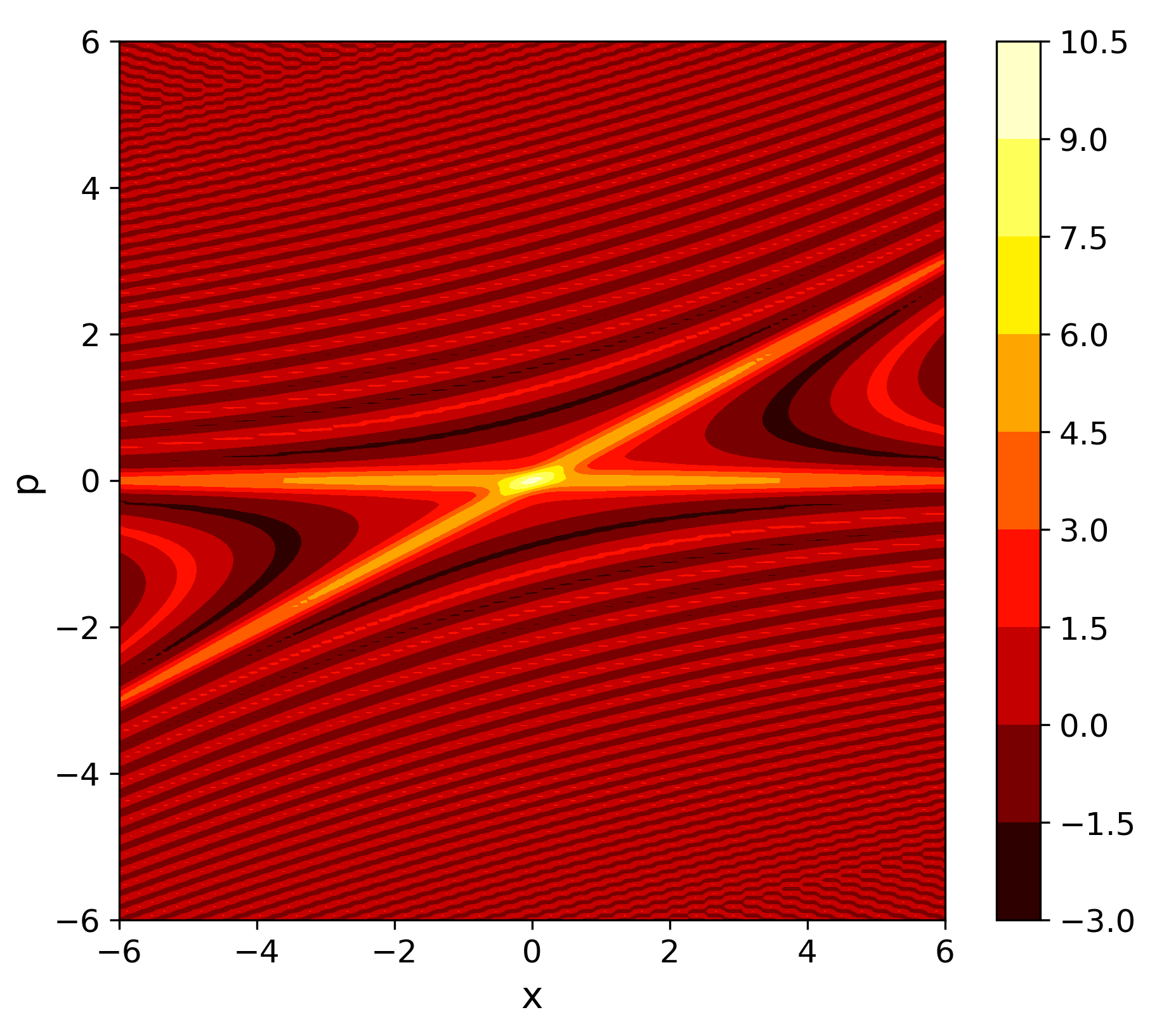}
            \end{figure}
            \vspace{-5mm}\centering (b2)
        \end{minipage}\hspace{1cm}
    \end{minipage}
\vspace{2mm}
\captionof{figure}{Same than Figure \ref{fig:incoh} but with $\alpha=\beta=\gamma=1$.\hfill \hspace{1mm}}
\label{fig:cohe}
\end{minipage}
\end{figure*}

The above expression lets us see how is the time evolution of the Wigner function, depending on the parameters $\alpha$, $\beta$ and $\gamma$ which are related with the coherence of the quantum system. When the system is a complete incoherent superposition, Fig. \ref{fig:incoh}, the only phase-space values in which Wigner function is different from zero are $p=0$ and $p=x/t$. However, when the system is a complete coherent superposition, Fig. \ref{fig:cohe}, an oscillating background appears. Its contour lines tend to be parallel to $p=0$ when $t\to\infty$, the same way as $W_{11}(x,p,t)$ does.

The presence of non-vanishing electric potential, makes a phase shift of the oscillation background\footnote{When $q\Delta\varphi\cdot\uptau=\pi/2+2\pi n$ with $n\in\mathbb{Z}$ and $t>\uptau$, the maximum contour lines of the oscillation background become minimum and vice versa.} that can be seen in Fig. \ref{fig:cohe}, so it changes the interference pattern when measuring. If we subtract the Wigner function with $\Delta\varphi\neq0$ minus the one with $\Delta\varphi=0$, then we obtain
\begin{eqnarray}\label{eq:dif_mixto}
    \Delta W(x,p,t)&=&\frac{4\rvert\gamma\rvert}{\sqrt{2\pi}}\left[\cos(\theta_{\Delta\varphi\neq0}(x,p,t))\right.\nonumber\\
    &&\left.-\cos(\theta_{\Delta\varphi=0}(x,p,t))\right].
\end{eqnarray}

By integrating $\Delta W(x,p,t)$ with respect to $x$ or to $p$, it is possible to find the difference between the interference pattern in the cases with electric device and without it. This difference between interference patterns is measurable, so the Aharonov-Bohm effect is thus again present.

From \eqref{eq:dif_mixto} we can see that a complete incoherent superposition of states ($\gamma=0$) does not present Aharonov effect. It shows that Aharonov-Bohm effect is a pure quantum phenomenon, i.e. it is not possible to observe that effect in a classical system.

\section{Conclusions}\label{sec5}

In this work, we have studied the Aharonov-Bohm effect within phase-space formalisms of non-relativistic quantum mechanism. The basic approach within these descriptions relies on quasiprobability distributions as the Wigner or Husimi functions, where it is not possible to understand the Aharonov-Bohm effect in terms of a measurable phase shift in the wave function of a charged particle. We have solved this problem by comparing directly different probabilities in the aforementioned phase-space formalisms. For such a purpose, we have first reviewed the basic properties of different phase-space approaches of quantum mechanics including the Segal-Bargmann space. We have explicitly shown how  canonical commutation relations are satisfied, which has taken us to review the inverse of the Weyl transform and the Moyal product. Once the bases have been set, we were able to derive the expected value for operators by using Wigner functions and the expressions describing time evolution of systems. On the other hand, we have derived the Segal-Bargmann space as a particular example of spaces $\mathcal{H} L^{2}(U, \alpha)$. Once it has been established, we have related it to quantum mechanics by means of commutation relations of creation and annihilation operators. At this point, we have also used canonical commutation relations indirectly.\\

By using these tools, we have posed the non-relativistic Aharonov-Bohm effect in phase space and Segal-Bargmann space. We begin by studying a particular case composed by a system in which the electric potential is not zero. By applying the proposed formalisms, quantifiable effects appear due to the presence of electric potential, even though electromagnetic fields are null within the region where the particles are propagated. Moreover, both formalisms predict the same results as the standard Schrödinger quantum mechanics formalism. Subsequently, we consider another case by studying the situation in which the magnetic potential is non-zero. By applying both formalisms to this system, we can find measurable effects owing to the presence of a magnetic vector potential, even though fields are null in the region accessible for the propagation of the particles. Conversely, for this system it is not possible to compute explicitly the probabilities of detecting a particle. For this reason, we compare the expressions obtained by means of Wigner and  Husimi quasiprobability distributions with the results in Schrödinger formalism by using the transforms that connect them. Similarly, the phase differences that we obtained in the corresponding formalisms agree with the expected ones. Subsequently, we pose a general system in which electric potential and magnetic vector potential are non-zero, in order to obtain a more general expression for the phase acquired by the wave function as a result of Aharonov-Bohm effect. This expression is expressed in a more compact fashion by using covariant notation, where it is easy to check that the effect is gauge independent.

At last, we elucidate the Aharonov-Bohm effect employing a density operator to characterize a superposition of states featuring position and momentum indeterminacy. The density operator is parameterized with real and complex coefficients, representing coherent and incoherent components. Temporal evolution is examined utilizing the Wigner function, giving raise to distinctive interference phenomena. The manifestation of the effect is illustrated through the temporal evolution of Wigner functions in the presence of an electric potential, discerning coherent and incoherent superpositions. The introduction of an electric potential induces a phase shift in the interference pattern, manifesting the Aharonov-Bohm effect. The analysis underscores that this phenomenon is intrinsically quantum mechanical, and its manifestation is absent in classical systems.

\section*{Acknowledgements}\label{Ack}

This work was partially supported by the MICINN (Spain) project PID2019-107394GB-I00/AEI/10.13039/501100011033 (AEI/FEDER, UE) and PID2022-139841NB-I00, COST (European Cooperation in Science and Technology) Actions CA21106 and CA21136. JARC acknowledges support by Institut Pascal at Université Paris-Saclay during the Paris-Saclay Astroparticle Symposium 2022, with the support of the P2IO Laboratory of Excellence (program “Investissements d’avenir” ANR-11-IDEX-0003-01 Paris-Saclay and ANR-10-LABX-0038), the P2I axis of the Graduate School of Physics of Université Paris-Saclay, as well as IJCLab, CEA, APPEC, IAS, OSUPS, and the IN2P3 master projet UCMN.”

\section*{Author Contribution Statement}\label{ACS}
All authors contributed equally to the study conception and design. Material
preparation, and analysis were performed by all authors. The first
draft of the manuscript was written cooperatively, and all authors commented
on previous versions of the manuscript. All authors read and approved
the final manuscript.

\section*{Data Availability Statement}\label{DAS}
Data sharing not applicable to this article as no datasets were generated
or analysed during the current study.
\bibliography{referencias}
\end{document}